\documentclass[preprint]{svjour3}
\smartqed  
\usepackage{amsmath}
\usepackage{amssymb}
\usepackage{bbm}

\usepackage{enumerate}
\usepackage{enumitem}
\usepackage{graphicx}
\graphicspath{{figures/}}

\usepackage{pgf}
\usepackage{tikz}
\usetikzlibrary{shapes.multipart, calc}
\usetikzlibrary{positioning,automata}
\tikzset{rect state/.style={draw,rectangle}}
\tikzset{sec state/.style={draw,rectangle,minimum size=0.9cm}}
\usepackage{multicol}

\usepackage{cite}

\usepackage{hyperref}

\usepackage{color}

\spnewtheorem{theorem1}{Theorem}{\bf}{\it}
\spnewtheorem{definition1}{Definition}{\bf}{\rm}
\spnewtheorem{observation}{Observation}{\bf}{\rm}
\spnewtheorem{assumption}{Assumption}{\bf}{\rm}
\spnewtheorem{remark1}{Remark}{\bf}{\rm}
\spnewtheorem{lemma1}[theorem]{Proposition}{\bf}{\rm}
\spnewtheorem{approach}{Approach}{\bf}{\rm}

\newcommand{\R}{R}

\newcommand{\G}{G}
\renewcommand{\H}{H}
\newcommand{\E}{E}
\newcommand{\e}{e}
\newcommand{\Q}{Q}
\newcommand{\q}{q}
\newcommand{\f}{f}

\newcommand{\A}{\mathcal{A}}
\newcommand{\X}{X}
\newcommand{\x}{x}

\newcommand{\lang}{\mathcal{L}}
\newcommand{\ioseq}{\mathcal{L}^{IO}}

\newcommand{\autotrans}{{T^{IO}}}

\newcommand{\obs}{\Theta}

\makeatletter
\providecommand*{\cupdot}{%
  \mathbin{%
    \mathpalette\@cupdot{}%
  }%
}
\newcommand*{\@cupdot}[2]{%
  \ooalign{%
    $\m@th#1\cup$\cr
    \sbox0{$#1\cup$}%
    \dimen@=\ht0 %
    \sbox0{$\m@th#1\cdot$}%
    \advance\dimen@ by -\ht0 %
    \dimen@=.5\dimen@
    \hidewidth\raise\dimen@\box0\hidewidth
  }%
}

\providecommand*{\bigcupdot}{%
  \mathop{%
    \vphantom{\bigcup}%
    \mathpalette\@bigcupdot{}%
  }%
}
\newcommand*{\@bigcupdot}[2]{%
  \ooalign{%
    $\m@th#1\bigcup$\cr
    \sbox0{$#1\bigcup$}%
    \dimen@=\ht0 %
    \advance\dimen@ by -\dp0 %
    \sbox0{\scalebox{2}{$\m@th#1\cdot$}}%
    \advance\dimen@ by -\ht0 %
    \dimen@=.5\dimen@
    \hidewidth\raise\dimen@\box0\hidewidth
  }%
}
\makeatother

\begin{document}

\title{A General Language-Based Framework for Specifying and Verifying Notions of Opacity
		\thanks{Research supported in part by US NSF under grants CNS-1738103, CNS-1801342, and ECCS-1553873.}}

\titlerunning{General Language-Based Opacity}        

\author{Andrew Wintenberg \and
        Matthew Blischke \and
		St\'{e}phane Lafortune \and
        Necmiye Ozay
}

\authorrunning{A. Wintenberg, M. Blischke, S. Lafortune, N. Ozay} 

\institute{A. Wintenberg \at
              \email{awintenb@umich.edu} \\
              M. Blischke\at
              \email{matblisc@umich.edu}\\
              S. Lafortune \at
              \email{stephane@umich.edu} \\
              N. Ozay \at
              \email{necmiye@umich.edu}   \\
              Department of EECS, University of Michigan,\\
1301 Beal Avenue, Ann Arbor, MI 48109-2122, USA        
}

\date{}

\maketitle

\begin{abstract}
Opacity is an information flow property that captures the notion of \textit{plausible deniability} in dynamic systems, that is whether an intruder can deduce that ``secret'' behavior has occurred.
In this paper we provide a general framework of opacity to unify the many existing notions of opacity that exist for discrete event systems.
We use this framework to discuss language-based and state-based notions of opacity over automata.
We present several methods for language-based opacity verification, and a general approach to transform state-based notions into language-based ones.
We demonstrate this approach for current-state and initial-state opacity, unifying existing results.
We then investigate the notions of $K$-step opacity.
We provide a language-based view of $K$-step opacity encompassing two existing notions and two new ones.
We then analyze the corresponding language-based verification methods both formally and with numerical examples.
In each case, the proposed methods offer significant reductions in runtime and space complexity.
\end{abstract}

\section{Introduction}
\label{sec:intro}
As modern systems become increasingly connected, information flow has become critical to their correct operation.
These systems have entered many areas of life in the form of autonomous vehicles, the smart grid, location-based services, and medical monitoring, to name but a few areas.
The increasing amount of physical and human interaction with these systems raises concerns over security and privacy.
Transmission of information across networks possesses an inherent risk of revealing private information to an outside observer called the \textit{intruder}, potentially with malicious intent.
Formal modeling of information flow properties has been proposed as a way to understand and manage these risks in networked dynamic systems.
Notions like non-interference \cite{focardiNonInterferenceAnalysis2000} and anonymity \cite{reiterCrowdsAnonymityWeb1998} were developed in the computer science community for this purpose.

More recently, the notion of \textit{opacity} \cite{mazareUsingUnificationOpacity2004} was proposed as a general information flow property capturing the notion of ``plausible deniability'':
opacity holds if an intruder cannot deduce sensitive information from their observations of a system's behavior.
Opacity was further developed for a variety of Discrete Event System (DES) models, including transition systems \cite{bryansOpacityGeneralisedTransition2008}, finite state automata \cite{sabooriNotionsSecurityOpacity2007}, Petri nets \cite{bryansModellingOpacityUsing2005}, timed automata \cite{cassezDarkSideTimed2009}, modular automata \cite{masopustComplexityDetectabilityOpacity2019}, and more.
Within these models, many notions of opacity have been proposed to capture different forms of private or secret information.
Of particular importance are language-based opacity \cite{linOpacityDiscreteEvent2011a}, current-state opacity \cite{sabooriNotionsSecurityOpacity2007}, initial-state opacity \cite{sabooriVerificationInitialstateOpacity2008}, and the related notions of $K$-step and infinite step opacity \cite{sabooriVerificationKstepOpacity2009a, SABOORI200946}.
In addition to the type of private information, the capabilities of the intruder are also integral to notions of opacity.
While many works in DES consider a single intruder  with static observations of observable events, more complex observation schemes have also been considered, such as decentralized observers in \cite{wuComparativeAnalysisRelated2013b} or dynamic observers in \cite{cassezDynamicObserversSynthesis2009}.
Opacity is an expressive notion of security.
Many existing security properties, including non-interference and anonymity, can be formulated as opacity \cite{hadjicostisIntroductionEstimationInference2020}.
Additionally, opacity has been utilized in practical applications, like the enforcement of privacy in language-based services \cite{wuEnsuringPrivacyLocationBased2014a}.
For a thorough review of works in opacity in the context of DES, as of 2016, please see \cite{jacobOverviewDiscreteEvent2016c}.

Although a variety of notions of opacity have been proposed, they may not directly capture the desired notion of privacy or security in a given networked system.
One approach to analyzing specific notions of opacity is to transform them into existing notions where existing methods can be applied.
While some transformations between the various forms of opacity over automata have been studied (for example between current-state, initial-state, language-based\cite{wuComparativeAnalysisRelated2013b}), it is unclear if other notions like $K$-step opacity are comparable or how to handle new notions.
The first contribution of this paper is to develop a systematic approach for specifying and analyzing various notions of opacity.
This is accomplished with a general definition of opacity extending the notion developed for transition systems \cite{bryansOpacityGeneralisedTransition2008}.
We use this framework to model language-based opacity over automata and present several methods for verification thereof.
Then we develop a general transformation between state-based and language-based notions of opacity.
Using this, state-based notions of opacity can be described by constructing automata to specify secret behavior and verified using language-based methods.
This approach is first demonstrated on the simple notions of current-state and initial-state opacity.
The resulting verification methods resemble the existing standard approaches for verification of these forms of opacity.

The second contribution of this paper is to apply the proposed framework and verification methods to the less well-understood notions of $K$-step and infinite step opacity.
Whereas current-state opacity only considers an intruder's current state estimate, $K$-step and infinite step opacity may involve the intruder $\textit{smoothing}$ their estimates, i.e., improving estimates of the past with current information.
While it may appear that these notions are incomparable \cite{yinNewApproachVerification2017}, we provide a unified view of two prominent existing notions of $K$-step opacity along with two new ones that emerge using our framework.
These notions are then transformed into language-based and hence current-state opacity.
Furthermore, the resulting language-based verification methods offers considerable advantages over existing methods.
We demonstrate this both formally and with numerical examples.

After some brief preliminaries, the remaining sections of this paper are organized as follows.
Section \ref{sec:general_opacity} presents a general behavioral definition of opacity.
Section \ref{sec:automata_opacity} discusses language-based and state-based opacity over automata and methods for verification.
Section \ref{sec:existing} applies these concepts to verifying current-state and initial-state opacity.
Section \ref{sec:k_step_opacity} defines $K$-step and infinite step opacity in relation to existing notions.
Section \ref{sec:verify_k_step} presents methods for verification of $K$-step opacity while Section \ref{sec:complexity_k_step}  discusses the complexity of these methods.
Section \ref{sec:inf_step} discusses verification of infinite step opacity.
Section \ref{sec:examples} presents numerical results comparing verification methods for $K$-step opacity.
Finally, Section \ref{sec:conclusion} concludes the paper.

\subsection{Preliminaries \& notation}
\label{sec:prelim}
We denote the natural numbers as $\mathbb{N} = \{0,1,2,\cdots\}$.
For a string $s\in E^*$, we denote the length of $s$ as $|s|$ and write $s=s_0\cdots s_{|s|-1}$.
A nondetermistic finite automaton (NFA) is defined by a tuple $\G = (\Q,\E, \f,\Q_0,\Q_m)$ with a finite set of states $Q$, events $\E$, transition function $f:Q \times E \rightarrow 2^Q$, initial states $\Q_0$ and marked states $\Q_m$.
Unless stated otherwise, the term automaton will refer to an  NFA.
We also extend $f$ to the domain $Q \times E^*$ in the standard way.
For arbitrary sets $\Q_0',\Q_m' \subseteq Q$, we define the language of $\G$ starting in $\Q_0'$ and marked by $\Q_m'$ as \begin{equation}
\lang_{\Q_m'}(\G, \Q_0') = \{ s \in E^* \mid \exists q_0 \in \Q_0' \ \exists q_m \in \Q_m' \ q_m \in \f(q_0, s) \} \, .
\end{equation}
Then the language generated by $\G$ is defined $\lang(\G) = \lang_{Q}(\G,\Q_0)$ and the language marked by $\G$ is defined $\lang_m(\G) = \lang_{\Q_m}(\G,\Q_0)$.
For automata $\G$ and $H$, we write $\G^c$ for the complement of $\G$, $\G^R$ for the reversal of $\G$, $\G \cdot \H$ for the concatenation of $\G$ and $H$, $\G \times \H$ for the product of $\G$ and $H$, and $\det(\G)$ for the determinization of $\G$ using the standard power set construction.
We also use the notation $P_{\E_o}:\E^* \rightarrow \E_o^*$ to denote the projection of strings with respect to observable events $\E_o \subseteq \E$.
These notions are defined in \cite{cassandrasIntroductionDiscreteEvent2008}.

\section{A general framework for opacity}
\label{sec:general_opacity}

In this section we present a general framework of opacity to formalize the intuition of a system having ``plausible deniability''.
In order to unify the different notions of opacity that exist for a variety of system models, we discuss systems in terms of their \textit{behavior}, taking the approach of \cite{willemsBehavioralApproachOpen2007}.
Consider a system under observation by an intruder.
We denote the set of possible behaviors or runs of the system as $\R$.
For example, $\R$ may be the set of solutions to a differential equation modeling a continuous-time system or $\R$ may be the language of an automaton modeling a discrete event system.
The intruder makes observations of this behavior in the space $O$ through an observation map $\obs:\R \rightarrow O$.
Opacity describes the inability of the intruder to discern a class of secret runs $\R_S\subseteq \R$ from a class of nonsecret runs $\R_{NS} \subseteq \R$.
This inability can either be total or partial.
\begin{definition1}
\label{def:total_opacity}
We say that $(\R_S,\R_{NS})$ is \emph{totally opaque} to $\obs$ if \begin{equation}
\obs(\R_S) \subseteq \obs(\R_{NS}) \, .
\end{equation}
\end{definition1}
\begin{definition1}
\label{def:partial_opacity}
We say that $(\R_S,\R_{NS})$ is \emph{partially opaque} to $\obs$ if \begin{equation}
\obs(\R_S) \cap \obs(\R_{NS}) \neq \emptyset \, .
\end{equation}
\end{definition1}
When the behavior is taken to be the runs of a transition system, total opacity corresponds to the notion of opacity given in \cite{bryansOpacityGeneralisedTransition2008}.

Specific notions of opacity correspond to different specifications of the secret and nonsecret runs of the system and capabilities of the intruder.
In this work, we focus on total opacity as it relates to desirable notions of privacy and security, e.g., all secrets are hidden.
Alternatively, partial opacity can express notions like \textit{diagnosability} \cite{linOpacityDiscreteEvent2011a}, e.g., faults can be detected.
While the secret and nonsecret behavior can be arbitrary sets, we often consider them to be complements.
That is to say that nonsecret behavior means behavior which is not secret $\R_S = \R \setminus \R_{NS}$.
In this case observe the following.
\begin{observation}
\label{obs:sec_complement}
If $\R_S = \R \setminus \R_{NS}$ then $(\R_S,\R_{NS})$ is totally opaque if and only if $(\R,\R_{NS})$ is totally opaque. This is because \begin{equation}
\obs(\R_S) \subseteq \obs(\R_{NS}) \ \Leftrightarrow \ \obs(\R) = \obs(\R_S) \cup \obs(\R_{NS}) \subseteq \obs(\R_{NS}) \, .
\end{equation}
So under this condition, it suffices to consider only $\R_{NS}$ and $\R$.
\hfill $\lozenge$
\end{observation}

\subsection{Joint \& separate opacity}
\label{sec:joint_separate}
More complex notions of privacy can involve multiple classes of possibly overlapping secret behaviors.
Consider a set of pairs of classes of secret and nonsecret behaviors $\{\R_S(i),\R_{NS}(i)\}_{i \in I}$ over an index set $I$.
We consider two forms of opacity over these pairs with respect to an observation map $\obs$.

\begin{definition1}
\label{def:joint_opacity}
We say that $\{\R_S(i),\R_{NS}(i)\}_{i \in I}$ is \emph{jointly opaque} to $\obs$ if \begin{equation}
\left(\bigcup_{i \in I} \R_S(i), \bigcap_{i\in I} \R_{NS}(i)\right) \text{ is totally opaque.}
\end{equation}
Joint opacity considers all secrets uniformly.
It requires that a run in one secret class can be explained by a run that is nonsecret in every class.
\end{definition1}

\begin{definition1}
\label{def:separate_opacity}
We say that $\{\R_S(i),\R_{NS}(i)\}_{i \in I}$ is \emph{separately opaque} to $\obs$ if \begin{equation}
\forall i \in I, \  (\R_S(i),\R_{NS}(i)) \text{ is totally opaque.}
\end{equation}
Separate opacity considers all secrets individually. 
It requires that a run in one secret class can be explained by a run that is nonsecret in that class, but perhaps secret in another class.
\end{definition1}

\begin{observation}
\label{obs:joint_sep}
When $|I| = 1$, joint and separate opacity reduce to total opacity.
When $|I| \geq 1$, joint opacity implies separate opacity.
For $I' \subseteq I$, joint (separate) opacity of $\{\R_{S}(i),\R_{NS}(i)\}_{i \in I}$ implies joint (separate) opacity of $\{\R_{S}(i),\R_{NS}(i)\}_{i \in I'}$, respectively.
\hfill $\lozenge$
\end{observation}

\section{Opacity over automata}
\label{sec:automata_opacity}
Automata are a widely used model in discrete event systems.
There are many existing notions of opacity for automata which capture different privacy and security properties.
We can express these notions in the framework presented in Section \ref{sec:general_opacity} as total opacity with appropriate choices of secret and nonsecret behavior and of the intruder.
When secret and nonsecret behaviors are given as languages marked by automata, we refer to this as \textit{language-based opacity}.
More generally, when secret and nonsecret behaviors are defined in terms of the automaton's events and properties of the states we refer to this as \textit{state-based opacity}.
For example, many state-based notions involve visits to states designated as secret or nonsecret.
It is known that some state-based notions of opacity like current-state and initial-state opacity can be efficiently transformed into language-based opacity as in \cite{wuComparativeAnalysisRelated2013b}.
In this section, we first discuss language-based opacity in the framework of Section \ref{sec:general_opacity} and present corresponding methods for verification.
We then develop a general transformation from state-based to language-based notions of behavior.
With this transformation, we describe how state-based opacity can be verified using language-based methods.

\subsection{Language-based opacity}
\label{sec:language_opacity}
Consider a finite automaton $\G = (\Q, \E, \f, \Q_0, \Q_m)$.
In the context of language-based opacity, the relevant behavior of $\G$ is simply the language it marks $\R = \lang_m(\G)$.
The state marking of $\G$ allows us to consider systems whose behaviors are not prefix-closed.
The secret and nonsecret behaviors of this system are given as sublanguages $\R_S,\R_{NS} \subseteq \R$.
We consider observations given by strings over an alphabet $\Gamma$ which corresponds to the space $O= \Gamma^*$ and an observation map $\obs:\R \rightarrow \Gamma^*$.
\begin{definition1}
\label{def:language_opacity}
Given an observation map $\obs$, we say $\G$ is \emph{language-based opaque} if $\obs(\R_S) \subseteq \obs(\R_{NS})$, or equivalently $(\R_S,\R_{NS})$ is totally opaque to $\obs$.
\end{definition1}
This definition corresponds to the notion of strong opacity in \cite{linOpacityDiscreteEvent2011a}.
When $\obs(\R_S)$ and $\obs(\R_{NS})$ are regular, language-based opacity is equivalent to a regular language containment which is well-understood.
To this end, in the context of language-based opacity we consider a setting where $\R_S$, $\R_{NS}$ are regular and $\obs$ preserves regularity.
For many existing notions of opacity, the nonsecret behavior of a system is simply the system's behavior that is not secret
\footnote{In fact we can always modify the behavior of the system to ensure this while preserving opacity properties.}.
So we consider when $\R_S = \R \setminus \R_{NS}$.
Additionally, as $\R_{NS}$ represents nonsecret behavior within $\R$, it is convenient to define it in terms of regular nonsecret specification language specifications $L_{NS}$ over $E$ so that $\R_{NS} = \R \cap L_{NS}$.
The language $L_{NS}$ is specified by an automaton $\H_{NS}$ such that $L_{NS} = \lang_m(\H_{NS})$ and so $\R_{NS} = \lang_m(\G \times \H_{NS})$.
Finally, we consider a class of observation maps $\obs$ that preserve regularity.
\begin{definition1}
\label{def:static_mask}
A \emph{static mask} over $\R$ is a mapping $\obs:\R \rightarrow \Gamma^*$ that satisfies \begin{enumerate}
\item $\obs(\epsilon) = \epsilon$,
\item $\forall s \in \R, \ \obs(s) = \obs(s_0)\cdots \obs(s_{|s|-1})$.
\end{enumerate}
Any function $\obs:\E \rightarrow \Gamma \cup \{\epsilon\}$ can be uniquely made into a static mask over $\R \subseteq E^*$ by concatenation.
Additionally, any valid composition of static masks is also a static mask.
\end{definition1}
For example, given a set of observable events $\E_o \subseteq \E$, the natural projection $P_{\E_o}$ is a static mask over $\R \subset \E^*$ with $\Gamma = \E_o$.
Given any automaton $\G=(\Q,\E, \f,\Q_0,\Q_m)$, we can construct an automaton that marks $\obs(\lang_m(\G))$ which in a slight abuse of notation we denote as $\obs(\G)$.
We construct $\obs(\G)$ by replacing the events of $\G$ with their observations under $\obs$.
Formally, $\obs(\G)$ is an automaton with $\epsilon$-transitions defined by $\obs(\G) = (\Q,\Gamma \cup \{\epsilon\}, f_{\obs},Q_0,\Q_m)$ where \begin{equation}
\forall \gamma \in \Gamma \cup \{\epsilon\} \quad f_{\obs}(q_1,\gamma) = \{q_2 \in Q \mid \exists e \in \E, \ \gamma=\obs(e), \ q_2 \in f(q_1,e)\}
\end{equation}
A similar construction is described in \cite{linOpacityDiscreteEvent2011a} for more general observation maps.
In this case when the observation mask $\obs$ is a static mask, $\obs(\R_S)$ and $\obs(\R_{NS})$ are regular languages.
Using Observation \ref{obs:sec_complement}, we see that language-based opacity of $G$ is equivalent to the regular language containment \begin{equation}
\label{eq:equiv_cont}
\obs(\R) \subseteq \obs(\R_{NS}), \text{ where } \obs(\R) = \lang_m(\obs(G)), \ \obs(\R_{NS}) = \obs(\G \times H_{NS}) \, .
\end{equation}

\subsection{Verification of language-based opacity}
\label{sec:verify_language_opacity}
By expressing language-based opacity as the well-studied problem of regular language containment, we can leverage existing techniques to verify opacity.
We present three methods to check this language containment.

As input, the following methods take an automaton $\G = (\Q, \E, \f, \Q_0, \Q_m)$ modeling the system, a nonsecret specification automaton $\H_{NS} = (\Q_{NS}, \E,$ $\f_{NS}, \Q_{NS,0}, \Q_{NS,m})$, and a static mask $\obs:\R \rightarrow \Gamma^*$ where $\R = \lang_m(\G)$.
These methods verify the total opacity of $(\R_S,\R_{NS})$ to $\obs$ where $\R_{NS} = \lang_m(\G \times \H_{NS})$ and $\R_S = \R \setminus \R_{NS}$.
This is done by verifying the equivalent containment of equation \eqref{eq:equiv_cont}.

\begin{approach}[Forward Comparison]
\label{app:forward_comparison}
A standard approach for verifying language containment utilizes the following equivalence: \begin{equation}
\obs(\R) \subseteq \obs(\R_{NS}) \ \Leftrightarrow \ \obs(\R) \cap \obs(\R_{NS})^c = \emptyset \, .
\end{equation}
We construct $\G_{FC} = \obs(\G) \times \det(\obs(\G \times \H_{NS}))^c$ so that $\lang_m(\G_{FC}) = \obs(\R) \cap \obs(\R_{NS})^c$.
Note determinization is required to construct the complement as $\obs(\G \times \H_{NS})$ is nondeterministic in general.
Hence $(\R_S,\R_{NS})$ is totally opaque if and only if $\G_{FC}$ marks the empty language.
We then verify opacity by ensuring $\G_{FC}$ contains no reachable, marked state.
\hfill $\lozenge$
\end{approach}

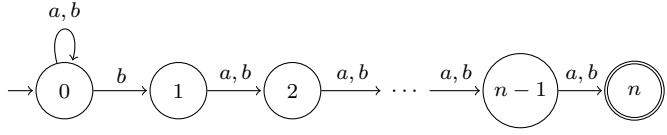
\begin{figure}
\centering
\begin{tikzpicture}[initial text=,shorten >=1pt,node distance=1.5cm,on grid, scale=1, transform shape, every text node part/.style={align=center}]
  \node[state,initial]   (0)              {$0$};
  \node[state]   (1) [right=of 0] {$1$};
  \node[state]   (2) [right=of 1] {$2$};
  \node   (3) [right=of 2] {$\cdots$};
  \node[state]   (4) [right=of 3] {$n-1$};
  \node[accepting, state]   (5) [right=of 4] {$n$};
  \path[->] (0) edge [loop above] node [above] {$a,b$} ()
				edge node [above] {$b$} (1)
			(1) edge node [above] {$a,b$} (2)
			(2) edge node [above] {$a,b$} (3)
			(3) edge node [above] {$a,b$} (4)
			(4) edge node [above] {$a,b$} (5);
\end{tikzpicture}
\caption{An automaton $\G_n$ with $n+1$ states. The forward determinization of $\det(\G_n)$ has $2^n+1$ states while the reverse determinization $\det(\G_n^R)$ has only $n+1$ states.}
\label{fig:exp_reverse}
\end{figure}

\begin{approach}[Reverse Comparison]
\label{app:reverse_comparison}
Instead of directly checking the language containment, recall that containment of languages is equivalent to containment of the reversed languages, therefore:
\begin{equation}
\obs(\R) \subseteq \obs(\R_{NS}) \ \Leftrightarrow \ \obs(\R)^R \subseteq \obs(\R_{NS})^R \, .
\end{equation}
Similar to the forward comparison method, we can construct $\G_{RC} = \obs(\G)^R \times \det(\obs(\G \times \H_{NS})^R)^c$ so that $\lang_m(\G_{RC}) = \obs(\R)^R \cap {(\obs(\R_{NS})^R)}^c$.
We then verify opacity by ensuring $\G_{RC}$ contains no reachable, marked state.
For some forms of opacity, reverse comparison significantly outperforms forward comparison.
This is possible because there are automata whose determinizations are exponentially larger than the determinizations of their reverses.
For example consider the automaton depicted in Figure \ref{fig:exp_reverse}.
\hfill $\lozenge$
\end{approach}

We can simplify the verification procedure by making assumptions on the structure of $\H_{NS}$.
Suppose that $\H_{NS}$ is a \textit{universal automaton}, i.e., $\lang(\H_{NS}) = E^*$.
Note that a given $\H_{NS}$ can be made to satisfy this by adding at most one state without effecting its marked language.
In this case $\G \times \H_{NS}$ will encode both $\R$ and $\R_{NS}$ with different sets of marked states.
With this observation, we can construct a deterministic finite automaton $\G_{SO} = \det(\obs(\G \times \H_{NS}))$ called the \textit{secret observer} which marks nonsecret observations.
With this automaton we can verify opacity using the following result.
\begin{lemma1}
\label{thm:secret_observer}
Suppose that $\lang(\H_{NS}) = E^*$.
Using the power set construction, define $\G_{SO} = (\overline{\Q}, \Gamma, \overline{\f}, \{\overline{\q}_0\}, \overline{\Q}_m)$ where $\overline{\Q} \subseteq 2^{\Q}$ so that $\G_{SO} = \det(\obs(\G \times \H_{NS}))$.
Then $(\R_S,\R_{NS})$ is totally opaque to $\obs$ if and only if for all $\gamma \in \lang(\G_{SO})$ it holds for $\overline{\q} = \overline{\f}(\overline{\q}_0, \gamma)$ that 
\begin{equation}
\label{eq:so_state_cond}
\overline{\q} \cap (\Q_m \times \Q_{NS}) = \emptyset \ \vee \ \overline{\q} \cap (\Q_m \times \Q_{NS,m}) \neq \emptyset \, .
\end{equation}
\end{lemma1}
\begin{proof}
First note that \begin{equation}
\lang(\G_{SO}) = \lang(\obs(\G)) \cap \lang(\obs(\H_{NS})) \supseteq \obs(\R) \cap \obs(\E^*) = \obs(\R) \, .
\end{equation}
For $\gamma \in \lang(\G_{SO})$ let $\overline{\q} = \overline{\f}(\overline{\q}_0, \gamma)$.
By the construction of $\G_{SO}$, note that
\begin{equation}
\exists \q \in \overline{\q} \cap \Q_m \times \Q_{NS} \ \Leftrightarrow \ \exists r \in \lang_m(\G) \cap \lang(H_{NS}) = \R \ \wedge \ \obs(r) = \gamma \, .
\end{equation}
Likewise, note that
\begin{equation}
\exists \q \in \overline{\q} \cap (\Q_m \times \Q_{NS,m}) \ \Leftrightarrow \ \exists r \in \lang_m(\G) \cap \lang_m(H_{NS}) = \R_{NS} \ \wedge \ \obs(r) = \gamma \, .
\end{equation}
Hence the state $\overline{\q}$ satisfies the conditions in \eqref{eq:so_state_cond} if and only if $\gamma \in \obs(\R_{NS})$ or $\gamma \not \in \obs(\R)$.
Combining these facts yields the result. \qed
\end{proof}
We use this result in the following approach.

\begin{approach}[Secret Observer]
\label{app:secret_observer}
Given $\lang(\H_{NS}) = E^*$, construct the secret observer $\G_{SO} = \det(\obs(\G \times \H_{NS}))$.
Using Proposition \ref{thm:secret_observer} we verify opacity by checking that every reachable state of $\G_{SO}$ satisfies the conditions in \eqref{eq:so_state_cond}.
\hfill $\lozenge$
\end{approach}

In each of these approaches, we verify opacity by constructing an automaton $\G_{FC}$, $G_{RC}$, or $G_{SO}$ and checking if each of its reachable states satisfies a given property.
As these are the largest automata constructed in these approaches, we quantify the complexity of these approaches in terms of the number of states in these automata.
We can improve these methods by incrementally constructing the reachable part of these automata and terminate if a violating state is found.

\begin{remark1}
\label{rem:secret_vs_forward}
When the secret observer method is applicable, i.e., $\lang(\H_{NS}) = \E^*$, the complexity of the secret observer method is always no worse than the complexity of the forward comparison method.
This is because both approaches require the construction of the automaton $\det(\obs(\G \times \H_{NS}))$, while this is all that is required for the secret observer method.
So for a given $\H_{NS}$ satisfying $\lang(\H_{NS}) = \E^*$, we do not consider the forward comparison method.
It is possible that a lower complexity could be obtained by a different choice of $\H_{NS}$ with $\lang(\H_{NS}) \neq \E^*$.
\hfill $\lozenge$
\end{remark1}

\subsection{Transforming state-based behavior}
\label{sec:transform_state}
We now discuss state-based notions of opacity in the framework of Section \ref{sec:general_opacity}.
Whereas in language-based opacity secret and nonsecret behaviors are defined solely in terms of the events, in  state-based opacity these behaviors are defined in terms of both events and properties of the states visited in the automaton.
As many existing notions of state-based opacity implicitly assume prefix-closed behavior, we consider a system modeled by an automaton $\A = (\X, \Sigma, \delta, \X_0)$ without marked states in the context of state-based opacity.
In order to express state-based opacity in the proposed framework of Section \ref{sec:general_opacity}, we must first identify the relevant behavior of the automaton.
While we could consider the behavior of the automaton to be state-event sequences, this description may contain more information than necessary.

\begin{figure}
\centering
\begin{tabular}{cc}
\begin{tikzpicture}[initial text=,shorten >=1pt,node distance=1.5cm,on grid, scale=1, transform shape, every text node part/.style={align=center}]
  \node[state,initial]   (0)              {$0$};
  \node[sec state]   (1) [right=of 0] {$1$};
  \node[state]   (2) [right=of 1] {$2$};
  \node[state]   (3) [below=of 1] {$3$};
  \node[sec state]   (4) [right=of 3] {$4$};
  \path[->] (0) edge node [above] {$\sigma_u$} (1)
				edge node [above] {$\sigma_o$} (3)
			(1) edge node [above] {$\sigma_o$} (2)
			(2) edge [loop above] node [above] {$\sigma_o$} ()
			(3) edge node [above] {$\sigma_u$} (4)
			(4) edge node [left] {$\sigma_o$} (2);
\end{tikzpicture}
&
\begin{tikzpicture}[initial text=,shorten >=1pt,node distance=2cm,on grid, scale=1, transform shape, every text node part/.style={align=center}]
  \node[state,initial]   (init)              {$\x_{init}$};
  \node[state,accepting]   (0) [above=of init] {$0$};
  \node[state,accepting]   (1) [right=of 0] {$1$};
  \node[state,accepting]   (2) [right=of 1] {$2$};
  \node[state,accepting]   (3) [below=of 1] {$3$};
  \node[state,accepting]   (4) [right=of 3] {$4$};
  \path[->] (init) edge node [left] {$(\sigma_{init},NS)$} (0)
            (0) edge node [above] {$(\sigma_u,S)$} (1)
				edge node [right] {$(\sigma_o,NS)$} (3)
			(1) edge node [above] {$(\sigma_o,NS)$} (2)
			(2) edge [loop above] node [above] {$(\sigma_o,NS)$} ()
			(3) edge node [above] {$(\sigma_u,S)$} (4)
			(4) edge node [left] {$(\sigma_o,NS)$} (2);
\end{tikzpicture}
\end{tabular}
\caption{On the left, an automaton $\A$ is depicted.
The labeling function $\ell$ is defined by labeling square states as $S$ and round states as $NS$.
On the right, the automaton $\G = \autotrans(\A,\ell)$ is depicted.
}
\label{fig:transform_ex}
\end{figure}
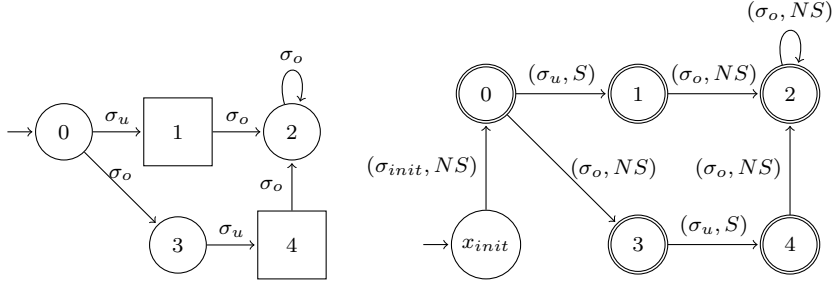

Consider when the secret behavior is defined by properties of the states rather than the states themselves.
We can model these properties with labels from a set $A$ assigned by the map $\ell:\X \rightarrow A$.
Viewing the events as inputs and state labels as outputs, we can describe the runs of this system as input-output sequences.
In this view, the system $(\A,\ell)$ is sometimes referred to as a \textit{Moore machine} \cite{cassandrasIntroductionDiscreteEvent2008}.
By writing these input-output sequences as sequences of pairs of an input (event) and the resulting output (state label), the resulting behavior is a language.
We introduce an artificial event $\sigma_{init}$ representing the system turning on to be paired with the label of the initial state.
The set of input-output sequences is then defined as follows.
\begin{definition1}
\label{def:io_seq}
Consider an automaton $\A = (\X,\Sigma, \delta,\X_0)$ with labeling map $\ell:\X \rightarrow A$.
Let $\sigma_{init}$ be disjoint from $\Sigma$ and define $\E = (\Sigma \cup \{\sigma_{init}\}) \times A$.
We define the \emph{input projection} $P^I:\Sigma^* \rightarrow (\Sigma \cup \{\sigma_{init}\})^*$ and \emph{output projection} $P^O:\Sigma^* \rightarrow A^*$ over input-output sequences by \begin{equation}
\begin{split}
\forall r = (\sigma_{0},a_0) \cdots (\sigma_{n-1},a_{n-1}) \in \E^*, \ \quad
& P^I(r) = \sigma_{0} \cdots \sigma_{n-1}, \\
& P^O(r) = a_0 \cdots a_{n-1} \, .
\end{split}
\end{equation}
Then set of \emph{input-output sequences of $\A$ under $\ell$} is defined as
\begin{align*}
\ioseq(\A, \ell) = \{ r \in E^+  \mid & \exists s \in \lang(\A) \ \exists \x_0 \cdots \x_{|r|-1} \in \X, \\
 & \x_0 \in \X_0, \ \forall i \in \{0, \cdots |r|-2\}, \ \x_{i+1} \in \delta(\x_i, s_i) \\
 &  P^I(r) = \sigma_{init} s_0 \cdots s_{|r|-2}, \ P^O(r) = \ell(\x_0) \cdots \ell(\x_{|r|-1}) \}
\end{align*}
\end{definition1}
We consider the behavior of $\A$ under $\ell$ to be $\R = \ioseq(\A,\ell)$.
For example consider the automaton $\A$ depicted in Figure \ref{fig:transform_ex}.
The run starting at state $0$ labeled $NS$, transitioning with event $\sigma_u$ to state $1$ labeled $S$, then transitioning with event $\sigma_o$ to state $2$ labeled $NS$, would be represented as $r=(\sigma_{init}, NS)(\sigma_{u},S)(\sigma_o,NS)$.

We can show that $\R$ is a regular language over $\E$ by constructing an automaton that marks it.
This is done by augmenting transitions in $\A$ with the state label of their destination.
Additionally, an artificial initial state $\x_{init}$ is introduced with transitions labeled with  $\sigma_{init}$ to the initial states of $\A$ to carry their state labels.
Formally, we define this transformation as follows.
\begin{definition1}
\label{def:state_transform}
Given $\A = (\X,\Sigma, \delta,\X_0)$ and $\ell:\X \rightarrow A$, define \emph{label-transform} of $\A$ by $\autotrans(\A,\ell) = (\Q,\E,\f,\Q_0, \Q_m)$, where $\Q = \X \cup \{\x_{init}\}$, $\E = (\Sigma \cup \{\sigma_{init}\}) \times A$, $\Q_0 = \{\x_{init}\}$, $\Q_m = \X$, and nonempty transitions defined by
\begin{equation}
\begin{split}
\forall a \in A, \quad & \f(\x_{init}, (\sigma_{init},a)) = \{\x_0 \in \X_0 \mid \ell(\x_0) = a\}, \\ 
\forall \x \in \X, \ \forall \sigma \in \Sigma, \ \forall a \in A, \quad & \f(\x,(\sigma,a)) = \{\x' \in \delta(\x,\sigma) \mid \ell(\x') = a\}
\end{split}
\end{equation}
\end{definition1}
An example of this transformation is depicted in Figure \ref{fig:transform_ex}.
Then by construction we have the following result.
\begin{lemma1}
\label{thm:transform}
Let $\A$ be an automaton with labeling map $\ell$.
Then the language marked by the label-transform of $\A$ is the same as the input-output sequence of $\A$. That is $\lang_m(\autotrans(\A,\ell)) = \ioseq(\A,\ell)$.
\end{lemma1}
In this way, we can transform the state-based behavior of one automaton into the language-based behavior of another.
We can use this label-transform to specify and verify state-based notions of opacity.

\subsection{Specification and verification of state-based opacity}
\label{sec:verify_state_opacity}

We can express state-based notions of opacity over an automaton $\A$ with state labels $\ell$ as total opacity over the input-output behavior $\R = \ioseq(\A, \ell)$.
Specifically, we consider the the total opacity of some secret and nonsecret behavior $\R_S,\R_{NS} \subseteq \R$ to an observation map $\obs:\R\rightarrow \Gamma^*$.
By modeling behavior in this way, where $\R$ is a regular language, we see that this state-based notion of opacity over $\A$ corresponds to a language-based one over $\G = \autotrans(\A,\ell)$.
Furthermore, under some simplifying assumptions, we can specify and verify state-based notions of opacity as in the language-based setting.
First we assume that the secret and nonsecret runs are specified as in the language-based setting.
\begin{assumption}
\label{asm:spec_auto}
There exists a nonsecret specification automaton $\H_{NS}$ such that \begin{equation}
L_{NS} = \lang_m(\H_{NS}), \quad \R_{NS} = \R \cap L_{NS}, \quad \R_S = \R \setminus \R_{NS} \, .
\end{equation}
\end{assumption}
Such specification automata $\H_{NS}$ for current-state and initial-state opacity are presented in Section \ref{sec:existing}.
\begin{remark1}
\label{rem:temporal_sec_spec}
Nonsecret behavior could also be specified with a temporal logic formula $\phi_{NS}$ with appropriate semantics.
From $\phi_{NS}$, the finite automaton $\H_{NS}$ marking runs that satisfy $\phi_{NS}$ could be synthesized.
In this way, opacity can be viewed as a temporal logic hyperproperty \cite{clarksonTemporalLogicsHyperproperties2014a}.
\hfill $\lozenge$
\end{remark1}
We will also require that $\obs$ is a static mask.
As the existing notions of opacity model observation as projection of strings with respect to a set observable events $\Sigma_o \subseteq \Sigma$, we only consider the induced observation map for simplicity.
By convention we will consider $\sigma_{init}$ to be observable, i.e., the intruder observes when the system turns on.
In this case we make the following assumption
\begin{assumption}
\label{asm:proj_obs_map}
The observation map of the intruder is induced by a set of observable events $\Sigma_o \subseteq \Sigma$.
This induced observation map is defined by $\obs:\Sigma^* \rightarrow \Gamma^*$ where $\Gamma = \Sigma_o \cup \{\sigma_{init}\}$ and $\obs = P_{\E_o \cup \{\sigma_{init}\}} \circ P^I$.
\end{assumption}
For an input-output pair $\sigma=(e,a) \in \Sigma$ the intruder observes $\obs(\sigma) = e$ for $e \in \E_o \cup \{\sigma_{init}\}$ and $\obs(\sigma) = \epsilon$ otherwise.
This observation map is a static mask as it is the composition of two static masks.

\begin{remark1}
\label{rem:state_obs_map}
Although not done here, one could also consider partially observable state outputs.
For example suppose that for each transition in $\A$, the intruder can see the event label of the transition if it is observable and also some function of the current state output $Y:A \rightarrow B$.
This corresponds to the observation map $\obs: \R \rightarrow ((\Sigma_o \cup \{\sigma_{init}\}) \times B)^*$ defined for $r \in \R$ with $t = P^I(r),\ a = P^O(r)$ by \begin{equation}
\obs(r) = (P_{\Sigma_o\cup\{\sigma_{init}\}}(t_0),Y(a_0)) \cdots (P_{\Sigma_o\cup\{\sigma_{init}\}}(t_{|a|-1}),Y(a_{|a|-1})) \, .
\end{equation}
In this setting, the intruder records an observation for each transition, even if the event label was unobservable.
We could similarly model the setting where the intruder would not know an unobservable event has occurred unless the state label changed with another construction.
\hfill $\lozenge$
\end{remark1}

Under these assumptions, notions of state-based opacity are specified by a nonsecret specification automaton $\H_{NS}$ and set of observable events $\Sigma_o$.
We will consider this setting in the remainder of this work.
We can then apply any of the language-based approaches of Section \ref{sec:verify_language_opacity} to $\G = \autotrans(\A,\ell)$, $\H_{NS}$, and the observation map $\obs$ induced by $\Sigma_{o}$ to verify the total opacity of $(\R_S, \R_{NS})$ to $\obs$.
This procedure is summarized in Figure \ref{fig:verification_pipeline}.
Due to the structure of $\G$ resulting from the transformation $\autotrans$, the secret observer method has a simple interpretation.
\begin{theorem1}
\label{thm:secret_obs_simp}
The pair $(\R_S, \R_{NS})$ is totally opaque to $\obs$ if every non-initial state of $\G_{SO} = \det(\obs(\G \times \H_{NS}))$ is marked.
\end{theorem1}
\begin{proof}
By construction, every non-initial state of $\G$ is marked.
Hence the conditions in \eqref{eq:so_state_cond} hold exactly when a secret observer state is the initial state or contains pair of states marked in $\G \times \H_{NS}$, i.e., the secret observer state is marked.
So by Proposition \ref{thm:secret_observer}, total opacity holds if every non-initial state of $\G_{SO}$ is marked.
\qed \end{proof}

\tikzstyle{input} = [rectangle, rounded corners, minimum width=3cm, minimum height=1cm,text centered, draw=black, fill=green!30]
\tikzstyle{process} = [rectangle, rounded corners, minimum width=2.5cm, minimum height=1cm, text centered, draw=black, fill=blue!30]
\tikzstyle{arrow} = [thick,->,>=stealth]

\begin{figure}
\centering

\begin{tikzpicture}[initial text=,scale=0.75, transform shape, node distance=3.5cm, every text node part/.style={align=center}]

\node (1) [input] {System Model: \\ Automaton $\A$ \\ State Labels $\ell$ \\ $\R = \ioseq(\A,\ell)$};
\node (2) [input, below= 0.75cm of 1] {Nonsecret Spec.: \\ Automaton $\H_{NS}$ \\ $L_{NS} = \lang_m(\H_{NS})$ \\ (CSO \& ISO - Sec. \ref{sec:existing} \\ $K$-step - Sec. \ref{sec:verify_k_step} \\ Inf. Step - Sec. \ref{sec:inf_step})};
\node (3) [input, below=0.75cm  of 2] {Observable Event \\ Set: $\Sigma_o \subseteq \Sigma$ };
\node (33) [process, right of=3] {Induced Obs. Map: \\ Static Mask $\obs$ \\ (Assumption \ref{asm:proj_obs_map})};
\node (4) [process, right of=1] {Transform \\ State Labels \\ $\G=\autotrans(\A, \ell)$ \\ (Sec. \ref{sec:transform_state})};
\node (5) [process, right of=2] {Construct \\ Nonsecret System \\ $\G_{NS}=\G \times \H_{NS}$};
\node (6) [process, right of=4] {Construct \\ Observations \\ $\obs(\G)$ \\ (Sec. \ref{sec:language_opacity})};
\node (7) [process, right of=5] {Construct \\ Observations \\ $\obs(\G_{NS})$ \\ (Sec. \ref{sec:language_opacity})};
\node (8) [process, above right= -3.3cm and 0.75cm of 6] {Verify Opacity as \\ Language Containment \\ $\lang_m(\obs(\G)) \subseteq \lang_m(\obs(\G_{NS}))$ \\ using approaches: \\ Forward Comparison \\ Reverse Comparison \\ Secret Observer \\ (Sec. \ref{sec:verify_language_opacity})};

\draw [arrow] (1) -- (4);
\draw [arrow] (2) -- (5);
\draw [arrow] (4) -- (5);
\draw [arrow] (4) -- (6);
\draw [arrow] (5) -- (7);
\draw [arrow] (6) -- (8);
\draw [arrow] (7) -- (8);
\draw [arrow] (3) -- (33);
\draw [arrow] (33) -| ($(6.west)-(5mm,5mm)$) |- ($(6.west)-(0,5mm)$);
\draw [arrow] (33) -| ($(7.west)-(5mm,5mm)$) |- ($(7.west)-(0,5mm)$);

\end{tikzpicture}
\caption{The proposed method for verifying state-based opacity by transforming to language-based opacity.
}
\label{fig:verification_pipeline}
\end{figure}
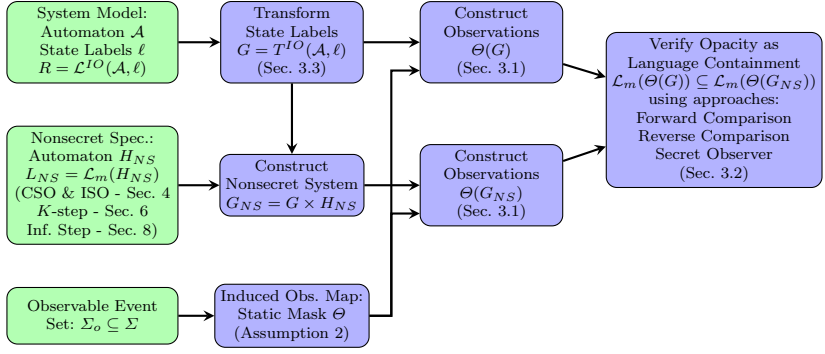
\vspace{-0.5cm}

\section{Current-state and initial-state opacity}
\label{sec:existing}
Several existing notions of opacity used in discrete event systems define secret behavior in terms of secret states of automata.
By viewing the secrecy of a state as a state output, we can describe the behavior of these systems as input-output sequences.
For a given state-based notion of opacity, we can identify the corresponding secret and nonsecret behavior in order to express this notion in our framework as total opacity.
By transforming the system, we can then apply any of the methods for verification of language-based opacity.
In particular we consider current-state opacity (CSO) and initial-state opacity (ISO).
While it is already known that CSO and ISO can be efficiently transformed into language-based opacity  \cite{wuComparativeAnalysisRelated2013b}, these examples demonstrate our transformation and provide insight into application to more complex state-based notions of opacity.

\subsection{Labeling secret states}
\label{sec:sec_states}
Consider an automaton $\A = (\X,\Sigma, \delta,\X_0)$ with a subset of states $\X_S \subseteq \X$ designated as secret and observable events $\Sigma_o \subseteq \Sigma$.
We also define the nonsecret states as $\X_{NS} = \X \setminus \X_S$.
This property of the states can be represented by labeling secret states with $S$ and other nonsecret states as $NS$.
So we define the set of labels $A = \{S,NS\}$ and labeling map $\ell:\X \rightarrow A$ by \begin{equation}
\ell(\x) = \begin{cases}
S, &  \x \in \X_S \\
NS, & \x \in \X_{NS}
\end{cases}
\end{equation}
A visit to a secret state in $\A$ following an event $\sigma \in \Sigma$ corresponds to the input-output pair $\e = (\sigma ,S)$.
Likewise starting in a secret state in $\A$ corresponds to the pair $\e = (\sigma_{init}, S)$.
Using this observation, we define the set of secret and nonsecret input-output pairs as \begin{equation}
\label{eq:sec_nonsec_pair}
\E_S = (\Sigma \cup \{\sigma_{init}\}) \times \{S\} \qquad \E_{NS} = (\Sigma \cup \{\sigma_{init}\}) \times \{NS\} \, .
\end{equation}
These sets can be used to specify the secret and nonsecret behavior in terms of the input-output sequences $\R = \ioseq(\A,\ell)$ for CSO and ISO.

\subsection{Current-state opacity (CSO)}
\label{sec:cso}
First we consider current-state opacity. 
Current state opacity describes the inability of an intruder to deduce that the current state of the system is secret.
It can be defined as follows.
\begin{definition1}[Current-State Opacity  \cite{falconeEnforcementValidationRuntime2015a}]
\label{def:cso}
An automaton $\A = (\X,\Sigma, \delta,\X_0)$ is said to be \emph{current-state opaque} with respect to the secret states $\X_S \subseteq \X$ and observable events $\Sigma_o \subseteq \Sigma$ if 
\begin{equation}
\begin{split}
&\forall \x_0 \in \X_0 \ \forall s \in \lang(\A) \ s.t. \ \exists \x_S \in \delta(\x_0,s) \cap \X_S, \\
&\exists \x_0' \in \X_0 \ \exists s' \in \lang(\A), \ P_{\Sigma_o}(s) = P_{\Sigma_o}(s')  \ \wedge \exists \x_{NS} \in \delta(\x_0',s') \cap \X \setminus \X_{S} \, .
\end{split}
\end{equation}
\end{definition1}

\begin{figure}[h!]
\centering
\begin{tikzpicture}[initial text=,shorten >=1pt,node distance=2cm,on grid, scale=1, transform shape, every text node part/.style={align=center}]
  \node[state,initial]   (0)                {$0$};
  \node[state,accepting]   (1)  [right=of 0]    {$1$};
  \path[->] (0) edge [bend left=30] node [above] {$\E_{NS}$} (1)
				edge [loop above] node [above] {$\E_S$} ()
			(1) edge [loop above] node [above] {$\E_{NS}$} ()
				edge [bend left=30] node [below] {$\E_{S}$} (0);
\end{tikzpicture}
\hspace{1cm}
\begin{tikzpicture}[initial text=,shorten >=1pt,node distance=2cm,on grid, scale=1, transform shape, every text node part/.style={align=center}]
  \node[state,initial]   (0)                {$0$};
  \node[state,accepting]   (1)  [right=of 0]    {$1$};
  \node[state]   (2)  [below=1.75cm of 1]    {$2$};
  \path[->] (0) edge node [above] {$\E_{NS}$} (1)
			    edge node [below] {$\E_{S}$} (2)
			(1)	edge [loop above] node [above] {$\E$} ()
			(2)	edge [loop above] node [above] {$\E$} ();
\end{tikzpicture}
\caption{The nonsecret specification automata $\H_{NS}$ for CSO (left) and ISO (right).}
\label{fig:cso_iso_spec}
\end{figure}
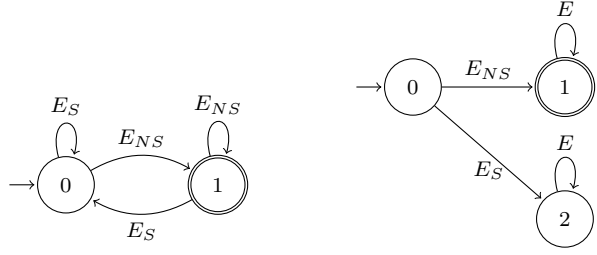

In words, runs of $\A$ ending with a visit to a secret state should look like a run ending with a visit to a nonsecret state.
In terms of input-output sequences, this definition divides the behavior $\R = \ioseq(\A,\ell)$ is into secret and nonsecret behavior $\R_S,\R_{NS} \subseteq \R$ defined by \begin{equation}
L_{NS} = \E^* \E_{NS}, \quad \R_{NS} = R \cap L_{NS}, \quad \R_{S} = R \setminus \R_{NS} \, ,
\end{equation}
where $\E_{NS}$ is defined in equation \eqref{eq:sec_nonsec_pair}.
We can use the nonsecret specification automaton $\H_{NS}$ depicted in Figure \ref{fig:cso_iso_spec} with $\lang_m(\H_{NS}) = L_{NS}$ so that Assumption \ref{asm:spec_auto} is satisfied.
Then using the observation map $\obs:\R \rightarrow \Gamma^*$ induced by the observable events $\Sigma_o$ as defined in Assumption \ref{asm:proj_obs_map}, we can see that $\A$ is current-state opaque if and only if $(\R_S,\R_{NS})$ is totally opaque with respect to $\obs$.
Hence we can use the language-based methods for verification.

\newpage

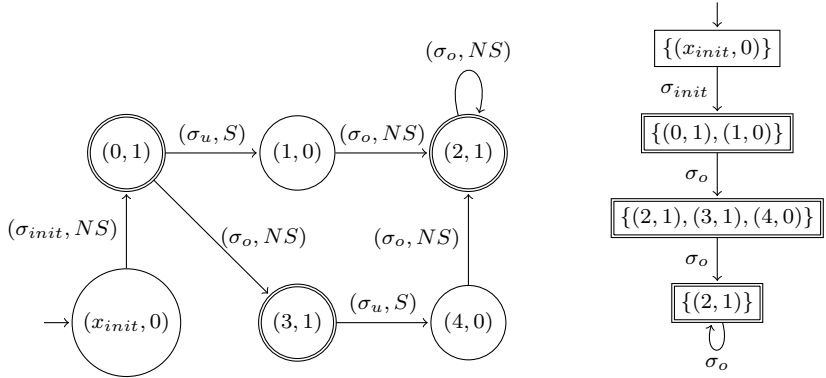
\begin{figure}
\centering
\begin{tikzpicture}[initial text=,shorten >=1pt,node distance=2.25cm,on grid, scale=1, transform shape, every text node part/.style={align=center}]
  \node[state,initial]   (init)              {$(x_{init},0)$};
  \node[state,accepting]   (0) [above=of init] {$(0,1)$};
  \node[state]   (1) [right=of 0] {$(1,0)$};
  \node[state,accepting]   (2) [right=of 1] {$(2,1)$};
  \node[state,accepting]   (3) [below=of 1] {$(3,1)$};
  \node[state]   (4) [right=of 3] {$(4,0)$};
  \path[->] (init) edge node [left] {$(\sigma_{init},NS)$} (0)
            (0) edge node [above] {$(\sigma_u,S)$} (1)
				edge node [right] {$(\sigma_o,NS)$} (3)
			(1) edge node [above] {$(\sigma_o,NS)$} (2)
			(2) edge [loop above] node [above] {$(\sigma_o,NS)$} ()
			(3) edge node [above] {$(\sigma_u,S)$} (4)
			(4) edge node [left] {$(\sigma_o,NS)$} (2);
\end{tikzpicture}
\hspace{1cm}
\begin{tikzpicture}[initial text=,shorten >=1pt,node distance=1.125cm,on grid, scale=1, transform shape, every text node part/.style={align=center}]
  \node[rect state,initial above]   (init)              {$\{(\x_{init},0)\}$};
  \node[rect state,accepting]   (0) [below=of init] {$\{(0,1),(1,0)\}$};
  \node[rect state,accepting]   (1) [below=of 0] {$\{(2,1),(3,1),(4,0)\}$};
  \node[rect state,accepting]   (2) [below=of 1] {$\{(2,1)\}$};
  \path[->] (init) edge node [left] {$\sigma_{init}$} (0)
            (0) edge node [left] {$\sigma_o$} (1)
            (1) edge node [left] {$\sigma_o$} (2)
            (2) edge [loop below] node [below] {$\sigma_o$} (1);
\end{tikzpicture}
\caption{The product $\G \times \H_{NS}$ (left) for $\G = \autotrans(\A,\ell)$ where $\A$ is from from Figure \ref{fig:transform_ex} and the nonsecret specification automaton $\H_{NS}$ for CSO from Figure \ref{fig:cso_iso_spec} and the corresponding secret observer $\G_{SO}$ (right).}
\label{fig:cso_ex}
\end{figure}

To do this we first construct $\G = \autotrans(\A, \ell)$. 
As $\lang(\H_{NS}) = \E^*$, using Theorem \ref{thm:secret_obs_simp} we can verify CSO of $\A$ by checking if every non-initial state of the secret observer $\G_{SO}=\det(\obs(\G \times \H_{NS}))$ is marked where $\G = \autotrans(\A, \ell)$.
As an example of this method, we verify the current-state opacity of $\A$ from Figure \ref{fig:transform_ex} using its transformation $\G = \autotrans(\A, \ell)$.
Assuming $\Sigma_o = \{\sigma_o\}$, we construct $\G \times \H_{NS}$ and $\G_{SO} = \det(\obs(\G \times \H_{NS}))$ which are depicted in Figure \ref{fig:cso_ex}.
As every non-initial state of $\G_{SO}$ is marked, we deduce $\A$ is CSO.

\begin{remark1}
\label{rem:label_movement}
The construction $\G = \autotrans(\A,\ell)$ essentially moves the state label information  from the states of $\A$ to the events of $\G$.
In the product $\G \times \H_{NS}$, these labels are then moved from the events back to the states in the form of state markings.
As a result $\G \times \H_{NS}$ is the same as the original automaton $\A$ where nonsecret states are marked and there are new initial states resulting from $\x_{init}$ in $\G$.
In this way the secret observer method is comparable to the standard method for verifying current-state opacity \cite{sabooriNotionsSecurityOpacity2007} which checks if each state of the observer of $\A$ contains a nonsecret state.
While our approach may seem convoluted for verifying CSO, the purpose of our discussion and of the above example are to demonstrate how our approach can be used to verify state-based notions of opacity in general.
\hfill $\lozenge$
\end{remark1}

\subsection{Initial-state opacity(ISO)}
\label{sec:iso}
Next, we discuss the notion of initial-state opacity.
Initial-state opacity describes the inability of an intruder to deduce that the initial-state of a run was secret.
It can be defined as follows.
\begin{definition1}[Initial-State Opacity \cite{wuComparativeAnalysisRelated2013b}]
\label{def:iso}
The automaton $\A=(\X,\Sigma,\delta,\X_0)$ is said to be \emph{initial-state opaque} with respect to secret states $\X_S \subseteq \X_0$  and observable events $\Sigma_o \subseteq \Sigma$ if
\begin{equation}
\begin{split}
&\forall \x_0 \in \X_S \ \forall s \in \lang(\A) \ s.t. \ \exists \x \in \delta(\x_0,s), \\
&\exists \x_0' \in \X_{NS} \ \exists s' \in \lang(\A), \ P_{\Sigma_o}(s) = P_{\Sigma_o}(s')  \ \wedge \ \exists x' \in \delta(\x_0',s') \, .
\end{split}
\end{equation}
\end{definition1}
Similar to the discussion of current-state opacity, we see that the initial-state opacity of $\A$ is equivalent to the total opacity of $(\R_S,\R_{NS})$ to the observation map $\obs$ induced by $\Sigma_o$ where \begin{equation}
L_{NS} = \E_{NS} \E^*, \quad \R_{NS} = R \cap L_{NS}, \quad \R_{S} = R \setminus \R_{NS} \, .
\end{equation}
We can construct $\H_{NS}$ as in Figure \ref{fig:cso_iso_spec} so that $\lang_m(\H_{NS}) = L_{NS}$ and $\lang(\H_{NS}) = \E^*$.
Applying the secret observer method in this case is similar to transforming initial-state opacity to current-state opacity as in \cite{wuComparativeAnalysisRelated2013b} and using the standard approach to verify current-state opacity. 

\section{K-step \& infinite step opacity}
\label{sec:k_step_opacity}
While current-state opacity captures the notion of hiding current secrets, $K$-step and infinite step opacity capture the notion of hiding past secrets.
In this section, we define state-based notions of $K$-step and infinite step opacity over automata.
We then show how these relate to the existing notions.

\subsection{State-based $K$-step opacity}
\label{sec:state_k_step}
Consider a system as described in Section \ref{sec:sec_states} consisting of an automaton $\A=(\X,\Sigma, \delta,\X_0)$ and map $\ell:\X \rightarrow A$ labeling secret states with behavior $\R = \ioseq(\A, \ell)$.
We are given a subset of observable events $\Sigma_o \subseteq \Sigma$ inducing the observation map $\obs$.
$K$-step opacity concerns visits to these secret states during the last $K$ observations made by the intruder.
We use the term \textit{observation epoch} to refer to the system's behavior between observations.
More specifically, the epoch starts when an observation is made and ends right before another observation is made or at the end of the run.
We consider two types of secret behavior that can be exhibited in an observation epoch.
In the first type, which we call \textbf{type 1}, \textit{at least one secret state is visited}.
In the second type, which we call \textbf{type 2}, \textit{only secret states are visited}.

In order to describe these observation epochs in terms of the input-output pairs $\E = (\Sigma \cup \{\sigma_{init}\}) \times A$, we define the sets of observable and unobservable input-output pairs by
\begin{equation}
\label{eq:observable_pair}
\E_o = \{\e \in \E \mid \obs(\e) \neq \epsilon\}, \quad \E_{uo} = \E \setminus \E_o \, .
\end{equation}
Here observability relates to the concepts of \textit{silent transitions} from \cite{hadjicostisIntroductionEstimationInference2020}.
An unobservable pair $\e \in \E_{uo}$ is silent in that $\obs(\e) = \epsilon$, while an observable pair $\e \in \E_o$ is not silent as $\obs(\e) \neq \epsilon$.
As we consider $\obs$ induced by the projection of observable events $\Sigma_o$, it holds that $\E_{o} = (\Sigma_o \cup \{\sigma_{init}\}) \times A$.
In order to describe the secrecy of an observation epoch, we use the previous definition of the sets of secret and nonsecret input-output pairs $\E_S,\E_{NS}$ as in equation \eqref{eq:sec_nonsec_pair}.
With this we make the following definition.
\begin{definition1}
\label{def:observation_epoch}
The set of \emph{observation epochs} is defined to be $L_{epoch} = \E_{o} \E_{uo}^*$.
The sets of \emph{observation epochs exhibiting type 1 or type 2 secrets}, respectively, are defined by
\begin{equation}
L_{epoch,S,1} = L_{epoch} \cap (\E^* \setminus \E^*_{NS}), \quad L_{epoch,S,2} =  L_{epoch} \cap \E_{S}^* \, .
\end{equation}
Likewise the sets of type 1 and type 2 nonsecret epochs are defined by
\begin{equation}
\begin{split}
L_{epoch,NS,1} &= L_{epoch} \setminus L_{epoch,S,1} = L_{epoch} \cap \E_{NS}^*, \\
L_{epoch,NS,2} &= L_{epoch} \setminus L_{epoch,S,2} = L_{epoch} \cap (\E^* \setminus \E_S^*) \, .
\end{split}
\end{equation}
\end{definition1}
Because every run in $\R$ starts with the input-output pair $(\sigma_{init},a)$ for some $a \in A$ and $(\sigma_{init},a) \in \E_o$ by definition, it holds that $\R \subseteq \E_o \E^* = L_{epoch}^+$.
This means any run $r \in \R$ can uniquely be written as a concatenation of observation epochs, i.e., $\exists M > 0$, $r = r_{epoch,0} \cdots r_{epoch,M-1}$ with $r_{epoch,i} \in L_{epoch}$ for all $i < M$.
We refer to the epoch $r_{epoch,M-k-1}$ as the epoch $k^{th}$ from the end or as $k$ epochs ago.
For $K$-step opacity, we define different classes of secret and nonsecret behavior for each epoch in the past, up to $K$ epochs ago.
For $k \leq K$ and type $j \in \{1,2\}$ secrets, we define
\begin{align}
L_{S,j}(k) &= L_{epoch}^* L_{epoch,S,j} L_{epoch}^k \, , \\
L_{NS,j}(k) &= L_{epoch}^+ \setminus L_{S,j}(k) =  (L_{epoch}^* L_{epoch,NS,j} L_{epoch}^k) \cup \bigcup_{i=1}^{k} L_{epoch}^i  \, .
\end{align}
We refer to $L_{S,j}(k)$ and $L_{NS,j}(k)$ as the \textit{$k$-delayed secret and nonsecret behavior} specifications, respectively.
Note that a run consisting of fewer than $k+1$ observation epochs is by definition not an element of $L_{S,j}(k)$ as a secret could not have occurred $k+1$ epochs ago.
The $k$-delayed secret and nonsecret behavior of $\R$ with type $j \in \{1,2\}$ secrets are then defined \begin{equation}
\label{eq:k_step_behavior}
\R_{S,j}(k) = R \cap L_{S,j}(k), \qquad \R_{NS,j}(k) = R \setminus \R_{S,j}(k) = R \cap L_{NS,j}(k) \, .
\end{equation}
By considering these secrets jointly, we can model an intruder deducing \textit{if} a secret occurred within $K$ epochs ago.
\begin{definition1}
\label{def:k_step_joint}
For  $K \in \mathbb{N} \cup \{\infty\}$, we say the system $\A$ with secrets labeled by $\ell$ is \emph{jointly $K$-step opaque with type $j$ secrets} if $\{(\R_{S,j}(k),\R_{NS,j}(k))\}_{k=0}^{K}$ as defined in \eqref{eq:k_step_behavior} is jointly opaque.
\end{definition1}
By considering these secrets separately, we can model an intruder deducing \textit{when} a secret occurred within $K$ epochs ago.
\begin{definition1}
\label{def:k_step_separate}
For  $K \in \mathbb{N} \cup \{\infty\}$, we say the system $\A$ with $\ell$ is \emph{separately $K$-step opaque with type $j$ secrets} if $\{(\R_{S,j}(k),\R_{NS,j}(k))\}_{k=0}^{K}$ as defined in \eqref{eq:k_step_behavior} is separately opaque.
\end{definition1}

For $K=\infty$ we refer to these definitions as \textit{infinite step opacity}.
While separate $K$-step opacity involves $\R_{NS,j}(k)$ and hence $L_{NS,j}(k)$ for $k \leq K$, joint opacity only involves their intersections.
For convenience we define for $K \in \mathbb{N} \cup \{\infty\}$ \begin{equation}
\begin{split}
L_{NS,j}^{joint}(K) =& \bigcap_{k=0}^K L_{NS,j}(k) = L_{epoch}^* L_{epoch,NS,j}^{K+1} \cup \bigcup_{k=1}^K L_{epoch,NS,j}^k \, ,
\end{split}
\end{equation}
so that $\bigcap_{k=0}^K \R_{NS,j}(k) = R \cap L_{NS,j}^{joint}(K)$.
In the joint sense, a run is secret if it consists entirely of nonsecret epochs or its last nonsecret epoch was at least $K+1$ epochs ago.

By comparing the nonsecret specification languages, we can relate the different notions of $K$-step opacity.
Because $L_{epoch,NS,1} \subseteq L_{epoch,NS,2}$, it holds that $L_{NS,1}(K) \subseteq L_{NS,2}(K)$ and thus $\R_{NS,1}(K) \subseteq \R_{NS,2}(K)$.
Hence joint and separate $K$-step opacity with type 1 secrets imply joint and separate $K$-step opacity with type 2 secrets, respectively.
Additionally using Observation \ref{obs:joint_sep}, we see that joint $K$-step opacity with type $j\in \{1,2\}$ secrets implies separate $K$-step opacity with type $j$ secrets.
These implications are depicted in Figure \ref{tab:opac_square}.
This figure also depicts the relation to the existing notions of $K$-step opacity derived in the next section.
Furthermore, joint and separate $(K+1)$-step opacity with type $j \in \{1,2\}$ secrets implies joint and separate $K$-step opacity with type $j$ secrets, respectively.

\tikzstyle{block} = [rectangle, minimum width=4cm, minimum height=1cm,text centered, draw=black, fill=white!30]
\begin{figure}[h!]
\centering
\begin{tikzpicture}[initial text=,scale=1, transform shape, node distance=1.5cm, every text node part/.style={align=center}]

\node[block]   (0)                {Joint, Type 1 \\ Strong \cite{falconeRuntimeEnforcementKstep2013}   \\ Trajectory-based  \cite{sabooriNotionsSecurityOpacity2007}};
\node[block, right=of 0]   (1) {Separate, Type 1 \\ New};
\node[block, below=1cm of 0]   (2) {Joint, Type 2 \\ New};
\node[block, right=of 2]   (3) {Separate, Type 2 \\ Weak  \cite{falconeRuntimeEnforcementKstep2013}   \\ Non-trajectory based  \cite{sabooriNotionsSecurityOpacity2007}};

\draw [arrow] (0) -- (1);
\draw [arrow] (0) -- (2);
\draw [arrow] (1) -- (3);
\draw [arrow] (2) -- (3);

\end{tikzpicture}

\caption{Types of $K$-step opacity.
Arrows indicate logical implication. For example, joint type 1 $K$-step opacity implies separate type 1 $K$-step opacity.}
\label{tab:opac_square}
\end{figure}
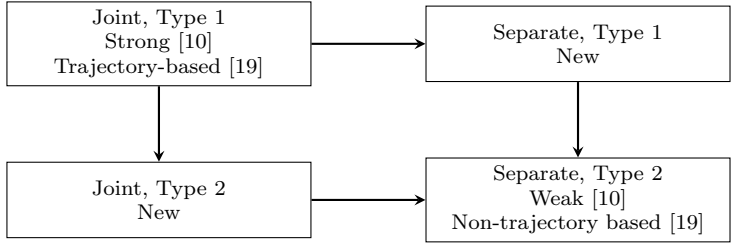

\subsection{Relation to existing notions of $K$-step opacity}
\label{sec:existing_k_step}
Now we show how these definitions relate to the existing notions of $K$-step opacity.
These notions were originally defined over deterministic finite automata, so for consistency we derive these relations in this setting.
Consider a deterministic automaton $\A=(\X,\Sigma, \delta,\{\x_o\})$ and interpret $\delta$ as a partial function $\delta:\X \times \Sigma \rightarrow \X$.
Consider a set of secret states $\X_S\subseteq \X$ and nonsecret states $\X_{NS} = \X \setminus \X_S$ as well as a set of observable events $\Sigma_o \subseteq \Sigma$.

The first form of $K$-step opacity was developed in \cite{sabooriNotionsSecurityOpacity2007}, and was later referred to as non-trajectory-based $K$-step opacity in \cite{SABOORI200946} and weak $K$-step opacity in \cite{falconeEnforcementValidationRuntime2015a}.
\begin{definition1}[$K$-step Weak Opacity \cite{falconeEnforcementValidationRuntime2015a}]
\label{def:weak_k_step}
The automaton $\A$ is \emph{weakly $K$-step opaque} with respect to $\X_S$ and $\Sigma_o$ if \begin{align*}
& (\forall uv \in \lang(\A) \ s.t. \ |P_{\Sigma_o}(v)| \leq K \ \wedge \delta(\x_0,u)\in \X_S ) \\
& (\exists u'v' \in \lang(\A)) \\
& (P_{\Sigma_o}(uv) = P_{\Sigma_o}(u'v') \ \wedge \ P_{\Sigma_o}(u) = P_{\Sigma_o}(u') \ \wedge \ 
\delta(\x_0,u') \in \X_{NS}) \, .
\end{align*}
\end{definition1}
The second version we consider is referred to as trajectory-based $K$-step opacity in \cite{SABOORI200946} and strong $K$-step opacity in \cite{falconeEnforcementValidationRuntime2015a}.
\begin{definition1}[$K$-step Strong Opacity \cite{falconeEnforcementValidationRuntime2015a}]
\label{def:strong_k_step}
The automaton $\A$ is \emph{strongly $K$-step opaque} with respect to $\X_S$ and $\Sigma_o$  if \begin{align*}
& (\forall t \in \lang(\A)) \\
& (\exists t' \in \lang(\A), \ \forall u',v' \ s.t. \ t'=u'v') \\
& (P_{\Sigma_o}(t) = P_{\Sigma_o}(t') \ \wedge  \ (|P_{\Sigma_o}(v')| \leq K \ \Rightarrow \ \delta(\x_0,u') \in \X_{NS})
\end{align*}
\end{definition1}

Weak $K$-step opacity describes the inability of the intruder to deduce an exact time of a visit to a secret state within the last $K$ observations.
Strong $K$-step opacity describes the inability of the intruder to deduce there was a visit to a secret state within the last $K$ observations.
With this intuition we can relate weak to separate and strong to joint opacity.
\begin{theorem1} 
\label{thm:k_step_equivalence}
Consider a deterministic automaton $\A$ with labeling map $\ell$ defined by the secret states $\X_S$ and observable events $\Sigma_o$.
Then
\begin{enumerate}
\item  Weak $K$-step opacity of $\A$ is equivalent to separate $K$-step opacity with type 2 secrets of $\A$.
\item  Strong $K$-step opacity of $\A$ is equivalent to joint $K$-step opacity with type 1 secrets of $\A$.
\end{enumerate}
\end{theorem1}
\begin{proof}
Because the automaton $\A$ is deterministic, there is a unique sequence of states associated with each string in $\lang(\A)$.
This defines a bijection $h:\R \rightarrow \lang(\A)$ where $\R = \ioseq(\A,\ell)$ such that \begin{equation}
\forall r \in R, \quad P^{I}(r) = \sigma_{init} \cdot h(r), \quad \obs(r) = \sigma_{init} \cdot P_{\Sigma_o}(h(r)) \, .
\end{equation}
Then note that we can write for $k \leq K$ 
\begin{multline}
h(\R_{NS,1}(k)) = \{ t \in \lang(\A) \mid \ \forall i \leq |t|, \ |P_{\Sigma_o}(t_i\cdots t_{|t|-1})|= k \ \Rightarrow \\ \delta(\x_0,t_0\cdots t_{i-1}) \in \X_{NS}\}
\end{multline}
\begin{multline}
h(\R_{NS,2}(k)) = \{ t \in \lang(\A) \mid \  |P_{\Sigma_o}(t)| < k \vee \exists i \leq |t| \ |P_{\Sigma_o}(t_i\cdots t_{|t|-1})|= k \ \wedge \\ \delta(\x_0,t_0\cdots t_{i-1}) \in \X_{NS})\} \, . 
\end{multline}

Suppose $\A$ is weakly $K$-step opaque and let $k \leq K$.
Consider a run of $\A$ given by $r \in \R$.
If $|\obs(r)| < k$ then by definition $r \in \R_{NS,2}(k)$.
Otherwise consider $t=h(r)$ so $|P_{\Sigma_o}(t)| \geq k$.
Let $i \leq |t|$ be such that $|P_{\Sigma_o}(t_{i}\cdots t_{|t|-1})| = k$ and define $u=t_0\cdots t_{i-1}$ and $v = t_i \cdots t_{|t|-1}$.
By weak opacity of $\A$, there must exist $t'=u'v'$ such that $P_{\Sigma_o}(t)=P_{\Sigma_o}(t')$, $|P_{\Sigma_o}(v')| = k$, and $\delta(\x_0,u') \in \X_{NS}$.
Thus for $r'=h^{-1}(t')$ it holds that  $r' \in \R_{NS,2}(k)$ and $\obs(r) = \obs(r')$.
Hence $\A$ is separately $K$-step opaque with type 2 secrets.
The proof of the converse is similar.

Now we consider strong $K$-step opacity.
Suppose that $\A$ is strongly $K$-step opaque.
Consider a run of $\A$ given by $r \in \R$ and define $t = h(r)$.
By strong $K$-step opacity of $\A$, there exists $t' \in \lang(\A)$ with $P_{\Sigma_o}(t)=P_{\Sigma_o}(t')$ where for every $i' \leq |t'|$ such that $|P_{\Sigma_o}(t'_{i'}\cdots t'_{|t'|-1})| \leq K$ it holds that $\delta(\x_0,t'_0\cdots t'_{i'-1}) \in \X_{NS}$.
Thus for $r'=h^{-1}(t')$ it holds that $r' \in \R_{NS,1}(k)$ for all $k \leq K$ and $\obs(r') = \obs(r)$.
Thus $\A$ is jointly $K$-step opaque with type 1 secrets.
The proof of the converse is similar.
\qed \end{proof}

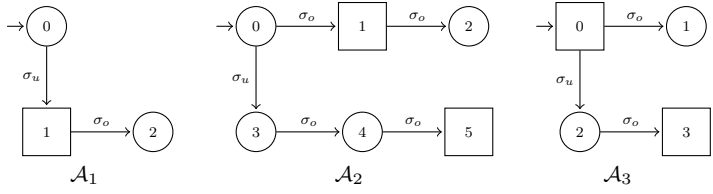
\begin{figure}[t!]
\centering
\begin{tabular}{c c c}
\begin{tikzpicture}[initial text=,shorten >=1pt,node distance=2cm,on grid, scale=0.7, transform shape, every text node part/.style={align=center}]
  \node[state,initial]   (0)                {$0$};
  \node[sec state]   (1) [below=of 0]  {$1$};
  \node[state]   (2) [right=of 1]  {$2$};
  \path[->] (0) edge node [left] {$\sigma_u$} (1)
  			(1) edge node [above] {$\sigma_o$} (2);
\end{tikzpicture} &
\begin{tikzpicture}[initial text=,shorten >=1pt,node distance=2cm,on grid, scale=0.7, transform shape, every text node part/.style={align=center}]
  \node[state, initial]   (0)                {$0$};
  \node[sec state]   (1) [right=of 0]  {$1$};
  \node[state]   (3) [below=of 0]  {$3$};
  \node[state]   (2) [right=of 1]  {$2$};
  \node[state]   (4) [right=of 3]  {$4$};
  \node[sec state]   (5) [right=of 4]  {$5$};
  \path[->] (0) edge node [above] {$\sigma_o$} (1)
  			    edge node [left] {$\sigma_u$} (3)
  			(1) edge node [above] {$\sigma_o$} (2)
  			(3) edge node [above] {$\sigma_o$} (4)
  			(4) edge node [above] {$\sigma_o$} (5);
\end{tikzpicture} & 
\begin{tikzpicture}[initial text=,shorten >=1pt,node distance=2cm,on grid, scale=0.7, transform shape, every text node part/.style={align=center}]
  \node[sec state,initial]   (0)                {$0$};
  \node[state]   (1) [right=of 0]  {$1$};
  \node[state]   (2) [below=of 0]  {$2$};
  \node[sec state]   (3) [right=of 2]  {$3$};
  \path[->] (0) edge node [above] {$\sigma_o$} (1)
				edge node [left] {$\sigma_u$} (2)
  			(2) edge node [above] {$\sigma_o$} (3);
\end{tikzpicture} \\
$\A_1$ & $\A_2$ & $\A_3$
\end{tabular}
\caption{Automata demonstrating the differences in the various notions of $K$-step opacity. Here square states denote secret states. The observable event set is $\Sigma_o = \{\sigma_o\}$.}
\label{fig:new_notion_ex_redo}
\end{figure}

The other notions of joint opacity with type 2 secrets and separate opacity with type 1 secrets, to our knowledge, have not been previously proposed.
The differences between the proposed notions of $K$-step opacity stem from how secrets interact with unobservable behavior.
To demonstrate how these notions differ, consider the automata $\A_1,\A_2,\A_3$ from Figure \ref{fig:new_notion_ex_redo} and secret states $X_S$ given by the square states and observable event set $\Sigma_o = \{\sigma_o\}$.
In $\A_1$ for example, there are no type 2 secret epochs possible as a visit to secret state $1$ must be preceded by a visit to nonsecret state $0$ in the same epoch.
Hence $\A_1$ is jointly and separately $1$-step opaque with type 2 secrets.
We can verify the various notions of $1$-step opacity for all of these automata as depicted in Table \ref{tab:new_notions}.

\begin{table}[h!]
\centering
\begin{tabular}{|c|c|c|c|c|}
\hline
1-Step Opacity Type & $\A_1$ & $\A_2$ & $\A_3$ \\ \hline
Separate Type 2 & Yes & Yes & Yes \\ \hline
Separate Type 1 & No & Yes & No \\ \hline
Joint Type 2 & Yes & No & No \\ \hline
Joint Type 1 & No & No & No \\ \hline
\end{tabular}
\caption{The results of verifying joint and separate $1$-step opacity with type 1 and type 2 secrets for the automata $\A_1,\A_2,\A_3$ from Figure \ref{fig:new_notion_ex_redo}.}
\label{tab:new_notions}
\end{table}

To paraphrase, joint $K$-step opacity with type 1 secrets reflects the inability of the intruder to deduce \textit{if} there was a period between observations where a \textit{single} secret state was visited, while joint $K$-step opacity with type 2 secrets reflects the inability of the intruder to deduce \textit{if} there was a period between observations where \textit{only} secret states were visited.
Likewise, separate $K$-step opacity with type 1 secrets reflects the inability of the intruder to deduce \textit{when} there was a period between observations where a \textit{single} secret state was visited, while separate $K$-step opacity with type 2 secrets reflects the inability of the intruder to deduce \textit{when} there was a period between observations where \textit{only} secret states were visited.
So we see for automata without unobservable events, type 1 and type 2 secrets are equivalent and these new notions of joint and separate reduce to the the existing notions of strong and weak.
While these new notions may only reflect differences in the modeling of unobservable events in some sense, they demonstrate how the proposed approach can be used to formulate precise notions of opacity appropriate for a given problem.

\section{Verification methods for finite K-step opacity}
\label{sec:verify_k_step}
In this section we will present methods for verification of $K$-step opacity for finite $K$.
First, we construct automata specifying nonsecret behavior.
Then we show how to use these automata to verify joint $K$-step opacity and separate $K$-step opacity.

\subsection{Nonsecret specification automata}
\label{sec:spec_k_step}
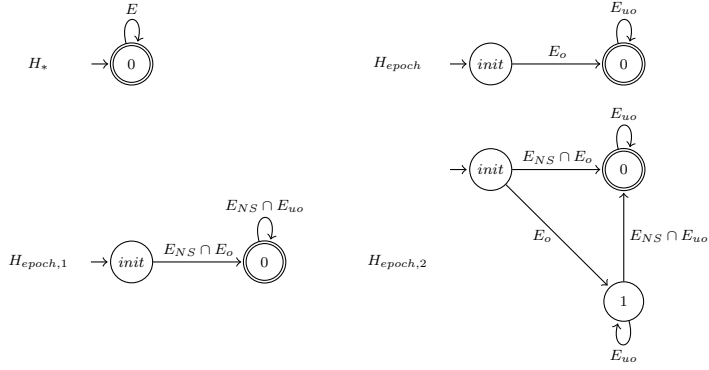
\begin{figure}[t]
\centering
\begin{tikzpicture}[initial text=,shorten >=1pt,node distance=2.5cm,on grid, scale=0.7, transform shape, every text node part/.style={align=center}]
 \node (hstar) {$\H_*$};
  \node[state,initial,accepting] [right=1.75cm of hstar] (init)                {$0$};
  \path[->] (init) edge [loop above] node [above] {$\E$} ();

\node (hepoch) [right=5cm of init] {$\H_{epoch}$};
  \node[state,initial]   (init_epoch)    [right=1.75cm of hepoch]            {$init$};
  \node[state,accepting] (0_epoch) [right=of init_epoch] {$0$};
  \path[->] (0_epoch) edge [loop above] node [above] {$\E_{uo}$} ()
		    (init_epoch) edge  node [above] {$\E_o$} (0_epoch);

\node (hepoch1) [below= 3.75cm of hstar] {$\H_{epoch,1}$};
  \node[state,initial]   (init_epoch_1)   [right=1.75cm of hepoch1]            {$init$};
  \node[state, accepting]   (0_epoch_1) [right=of init_epoch_1]  {$0$};
  \path[->] (init_epoch_1) edge node [above] {$\E_{NS} \cap \E_o$} (0_epoch_1)
  			(0_epoch_1) edge [loop above] node [above] {$\E_{NS} \cap \E_{uo}$} ();

\node (hepoch2) [below=3.75cm of hepoch] {$\H_{epoch,2}$};
  \node[state,initial]   (init_epoch_2) [above right= 1.75cm and 1.75cm of hepoch2]               {$init$};
  \node[state, accepting]   (0_epoch_2) [right=of init_epoch_2]  {$0$};
  \node[state]   (1_epoch_2) [below=of 0_epoch_2]  {$1$};
  \path[->] (init_epoch_2) edge node [above] {$\E_{NS} \cap \E_o$} (0_epoch_2)
				   edge node [left] {$\E_o$} (1_epoch_2)
  			(0_epoch_2) edge [loop above] node [above] {$\E_{uo}$} ()
  			(1_epoch_2) edge [loop below] node [below] {$\E_{uo}$} ()
  			    edge node [right] {$\E_{NS} \cap \E_{uo}$} (0_epoch_2);
\end{tikzpicture}
\caption{Automata used to construct nonsecret specification automata for $K$-step opacity defined over input-output pairs $\E$ categorized into nonsecret pairs $\E_{NS}$ as defined in equation \eqref{eq:sec_nonsec_pair}, and observable and unobservable pairs $\E_{o},\E_{uo}$ as defined in equation \eqref{eq:observable_pair}.
}
\label{fig:epoch_auto}
\end{figure}

In order to use language-based methods to verify $K$-step opacity, we must first construct automata that mark the corresponding nonsecret specification languages.
To do this, we will use the automata depicted in Figure \ref{fig:epoch_auto} as building blocks.
These automata are defined in terms of the input-output pairs $\E$ categorized into nonsecret pairs $\E_{NS}$ as defined in equation \eqref{eq:sec_nonsec_pair}, and observable and unobservable pairs $\E_{o},\E_{uo}$ as defined in equation \eqref{eq:observable_pair}.
Note that $\lang_m(\H_*) = \E^*$, $\lang_m(\H_{epoch}) = L_{epoch}$, $\lang_m(\H_{epoch,1}) = L_{epoch,NS,1}$, and $\lang_m(\H_{epoch,2}) = L_{epoch,NS,2}$
\footnote{
While $\H_{epoch,2}$ could be designed to be deterministic, our nondeterministic $\H_{epoch,2}$ offers reduced complexity.
}.
To construct automata that specify the nonsecret runs using these building blocks, we introduce the following notation.
\begin{definition1}
\label{def:concatenation}
Let $H^i = (\Q^i,\E,\f^i,\Q_0^i,\Q_m^i)$ for $i \in \{1,2\}$ be such that $\Q^2_0 \cap \Q^2_m = \emptyset$.
Let $\Q^{\cupdot} = \Q^1 \sqcup \Q^2 \setminus \Q^2_0$, $\Q^{\cupdot}_0 = \Q^1_0 \cup \Q^1_m$, and $\Q^{\cupdot}_m = \Q^2_m$.
Here $\sqcup$ denotes the disjoint union.
We define the \emph{concatenated automaton} $H^1 \cupdot H^2 = (\Q^{\cupdot}, \E, \f^{\cupdot}, \Q^{\cupdot}_0, \Q^{\cupdot}_m)$ where for all $\sigma \in \E$,\begin{align*}
\forall \q^1 \in \Q^1 \setminus \Q^1_m, \ \f^{\cupdot}(\q^1,\sigma) &= \f^1(\q^1,\sigma) \\
\forall \q^2 \in \Q^2 \setminus \Q^2_0, \ \f^{\cupdot}(\q^2,\sigma) &= \f^2(\q^2,\sigma) \\
\forall \q^1 \in \Q_m^1, \ \f^{\cupdot}(\q^1, \sigma) &=  \f^1(\q^1,\sigma) \cup \bigcup_{\q^2 \in \Q^2_0} \f^2(\q^2,\sigma) \, .
\end{align*}
This construction merges the marked states of $H^1$ with the initial states of $H^2$.
Note that $\lang_m(\H^1 \cupdot H^2) = (\lang_m(\H^1) \cup \lang_{mm}(\H^1)) \cdot \lang_m(\H^2)$, where $\lang_{mm}(\H^1) = \lang_m(\H^1,\Q^1_m)$ is the marked language of $H^1$ starting at the marked states of $H^1$.
\end{definition1}
With this we can define suitable nonsecret specification automata by concatenating the automata from Figure \ref{fig:epoch_auto}.
\begin{definition1}
\label{def:spec_auto}
We define the \emph{nonsecret specification automata for $K$-step opacity} iteratively as follows.
Let $\H_{NS,j}(0)=\H_{NS,j}^{joint}(0)=\H_* \cupdot \H_{epoch,j}$ and for $k \geq 0$ define 
\begin{equation}
{\H}_{NS,j}(k+1) = {\H}_{NS,j}(k) \cupdot \H_{epoch}, \qquad
{\H}_{NS,j}^{joint}(k+1) = {\H}_{NS,j}^{joint}(k) \cupdot \H_{epoch,j}  \, .
\end{equation}
\end{definition1}
The following result relates these nonsecret specification automata to the $K$-delayed nonsecret behavior defining $K$-step opacity.

\begin{lemma1}
\label{thm:spec_auto}
For every $K \in \mathbb{N}$ it holds that\begin{equation}
\begin{split}
L_{epoch}^+ \cap \lang_m({\H}_{NS,j}(K)) &= L_{NS,j}(K), \\
L_{epoch}^+ \cap \lang_m({\H}^{joint}_{NS,j}(K)) &= L_{NS,j}^{joint}(K) \, .
\end{split}
\end{equation}
\end{lemma1}
\begin{proof}
We show this for $\H_{NS,2}(K)$.
The proofs for the other cases are similar.
We claim that for all $k \leq K$ that $\lang_{mm}(\H_{NS,2}(k)) = \E_{uo}^*$ and \begin{equation}
\label{eq:ind_hyp}
\lang_m(\H_{NS,2}(k)) = \E^* L_{epoch,NS,2} L_{epoch}^k \cup \E_{uo}^* \bigcup_{i=1}^{k} L_{epoch}^{i} \, .
\end{equation}
Note that this condition holds for $k=0$ as \begin{equation}
\lang_m(\H_{NS,2}(0)) = \E^* L_{epoch,NS,2}, \qquad \lang_{mm}(\H_{NS,2}(0)) = \E_{uo}^* \, .
\end{equation}
Now assume that condition \eqref{eq:ind_hyp} holds for some $k < K$.
Then by definition of $\cupdot$ we have \begin{align*}
\lang_m(\H_{NS,2}(k+1)) &=
 (\lang_m(\H_{NS,2}(k))\cup \lang_{mm}(\H_{NS,2}(k))) \lang_m(\H_{epoch}) \\
 &= (\E^* L_{epoch,NS,2} L_{epoch}^k \cup \E_{uo}^* \cdot \bigcup_{i=1}^{k} L_{epoch}^{i} \cup \E_{uo}^*) L_{epoch} \\
 & = \E^* L_{epoch,NS,2} L_{epoch}^{k+1} \cup \E_{uo}^* \cdot \bigcup_{i=1}^{k+1} L_{epoch}^{i} 
\end{align*}
Hence by induction, condition \eqref{eq:ind_hyp} holds for all $k \leq K$.
Then note because $L_{epoch} = \E_o \E_{uo}^*$ that \begin{align*}
L_{epoch}^+ \cap \lang_m(\H_{NS,2}(k))
&= L_{epoch}^* L_{epoch,NS,2} L_{epoch}^k \cup \bigcup_{i=1}^k L_{epoch}^i \\
&= L_{NS,2}(k) \, .
\end{align*}
\qed \end{proof}

We can then use the automata $\H_{NS,j}(K)$ and $\H_{NS,j}^{joint}(K)$ in specifying $K$-step opacity.
As before, consider an automaton $\A$ with secret states labeled by $\ell$ with behavior given by the input-output pairs $\R = \ioseq(\A, \ell)$.
Then in terms of equation \eqref{eq:k_step_behavior},
\begin{equation}
 R \cap L_{NS,j}(K) = \R_{NS,j}(K), \qquad
R \cap L_{NS,j}^{joint}(K) = \bigcap_{k=0}^K \R_{NS,j}(k) \, .
\end{equation}
So $\H_{NS,j}(k)$ for $k \leq K$ can be used as nonsecret specification automata for verification of separate $K$-step opacity with type $j$ secrets.
Likewise, $\H_{NS,j}^{joint}(K)$  can be used for joint $K$-step opacity with type $j$ secrets.
In any case, it holds that $\lang(\H_{NS,j}(K))=\lang(\H_{NS,j}^{joint}(K)) = \E^*$ so we will be able to apply the secret observer method later on.

\begin{remark1}
\label{rem:naming_conv}
By expanding the recursive definitions of $\H=\H_{NS,j}(K)$ or $\H=\H_{NS,j}^{joint}(K)$, we can write $\H$ in the form $\bigcupdot_{i=0}^{K+1} \H^i$.
To avoid ambiguity due to redundant state names, we refer to the state $\q$ of $\H^i$ by $(\q,i)$ when embedded in $\H$.
\hfill $\lozenge$
\end{remark1}

\subsection{Verification of joint $K$-step opacity}
\label{sec:verify_joint}
Using the nonsecret specification automaton $\H_{NS,j}^{joint}(K)$ for $K \in \mathbb{N}$, we can verify joint $K$-step opacity with type $j$ secrets as follows.
\begin{approach}[Joint $K$-step opacity verification]
\label{app:joint_ver}
Given $\A$, $\ell$, $\Sigma_o$, and $K < \infty$, construct  the label-transform $\G = \autotrans(\A, \ell)$, the nonsecret specification automaton $\H_{NS,j}^{joint}(K)$, and the static mask $\obs$ induced by $\Sigma_o$.
We can then apply any of the language-based methods from Section \ref{sec:verify_language_opacity} to $\G,\H_{NS,j}^{joint}(K)$, and $\obs$ to verify the joint $K$-step opacity with type $j$ secrets of $\A$.
\hfill $\lozenge$
\end{approach}

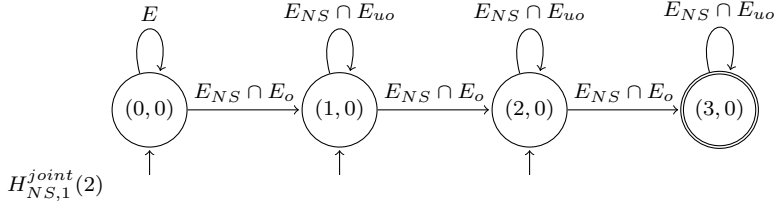
\begin{figure}
\centering
$\H_{NS,1}^{joint}(2)$
\begin{tikzpicture}[initial text=,shorten >=1pt,node distance=2.5cm,on grid, scale=1, transform shape, every text node part/.style={align=center}]
  \node[state,initial below]   (init)                {$(0,0)$};
  \node[state,initial below]   (0)  [right=of init]    {$(1,0)$};
  \node[state,initial below]   (1)  [right=of 0]    {$(2,0)$};
  \node[state,accepting]   (2)  [right=of 1]    {$(3,0)$};
  \path[->] (init) edge [loop above] node [above] {$\E$} ()
  				   edge node [above] {$\E_{NS} \cap \E_o$} (0)
  			(0)	   edge node [above] {$\E_{NS} \cap \E_o$} (1)
  				   edge [loop above] node [above] {$\E_{NS} \cap \E_{uo}$} ()
  			(1)   edge [loop above] node [above] {$\E_{NS} \cap \E_{uo}$} ()
 			      edge node [above] {$\E_{NS} \cap \E_o$} (2)
  			(2) edge [loop above] node {$\E_{NS} \cap \E_{uo}$} ();
  \end{tikzpicture}
\caption{The nonsecret specification automaton $\H_{NS,1}^{joint}(2)$ for 2-step joint opacity with type 1 secrets.}
\label{fig:joint}
\end{figure}

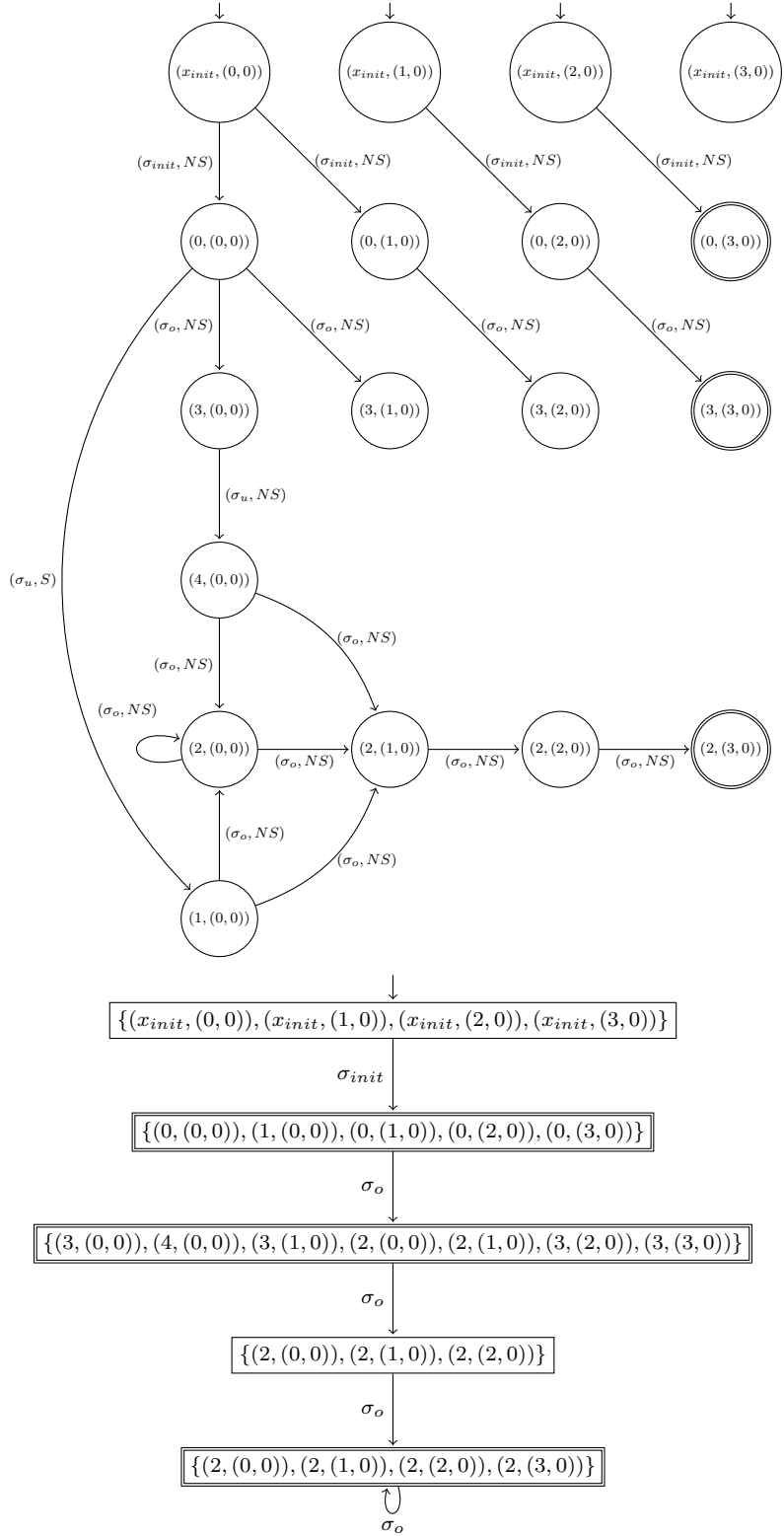
\begin{figure}
\centering
\begin{tikzpicture}[initial text=,shorten >=1pt,node distance=3.25cm,on grid, scale=0.7, transform shape, every text node part/.style={align=center}]
  \node[state,initial above]   (i0)                {$(\x_{init},(0,0))$};
  \node[state,initial above]   (i1) [right=of i0] {$(\x_{init},(1,0))$};
  \node[state,initial above]   (i2) [right=of i1] {$(\x_{init},(2,0))$};
  \node[state,initial above]   (i3) [right=of i2] {$(\x_{init},(3,0))$};
  \node[state]   (00) [below=of i0] {$(0,(0,0))$};
  \node[state]   (01) [right=of 00] {$(0,(1,0))$};
  \node[state]   (02) [right=of 01] {$(0,(2,0))$};
  \node[state,accepting]   (03) [right=of 02] {$(0,(3,0))$};
  \node[state]   (30) [below=of 00] {$(3,(0,0))$};
  \node[state]   (31) [right=of 30] {$(3,(1,0))$};
  \node[state]   (32) [right=of 31] {$(3,(2,0))$};
  \node[state,accepting]   (33) [right=of 32] {$(3,(3,0))$};
  \node[state]   (40) [below=of 30] {$(4,(0,0))$};
  \node[state]   (20) [below=of 40] {$(2,(0,0))$};
  \node[state]   (21) [right=of 20] {$(2,(1,0))$};
  \node[state]   (22) [right=of 21] {$(2,(2,0))$};
  \node[state,accepting]   (23) [right=of 22] {$(2,(3,0))$};
  \node[state]   (10) [below=of 20] {$(1,(0,0))$};
  
  \path[->] (i0) edge node [left] {$(\sigma_{init},NS)$} (00)
		         edge node [right] {$(\sigma_{init},NS)$} (01)
		    (i1) edge node [right] {$(\sigma_{init},NS)$} (02)
		    (i2) edge node [right] {$(\sigma_{init},NS)$} (03)
		    (00) edge [bend right=45] node [left] {$(\sigma_u,S)$} (10)
		         edge node [left] {$(\sigma_o,NS)$} (30)
		         edge node [right] {$(\sigma_o,NS)$} (31)
            (01) edge node [right] {$(\sigma_o,NS)$} (32)
            (02) edge node [right] {$(\sigma_o,NS)$} (33)
            (10) edge node [right] {$(\sigma_o,NS)$} (20)
                 edge [bend right=25] node [right] {$(\sigma_o,NS)$} (21)
            (20) edge [loop left] node [above left= 5mm and -5mm] {$(\sigma_o,NS)$} (20)
                 edge node [below] {$(\sigma_o,NS)$} (21)
            (21) edge node [below] {$(\sigma_o,NS)$} (22)
            (22) edge node [below] {$(\sigma_o,NS)$} (23)
            (30) edge node [right] {$(\sigma_u,NS)$} (40)
            (40) edge [bend right=0] node [left] {$(\sigma_o,NS)$} (20)
                 edge [bend left=25] node [right] {$(\sigma_o,NS)$} (21);
  \end{tikzpicture}
\begin{tikzpicture}[initial text=,shorten >=1pt,node distance=1.5cm,on grid, scale=1, transform shape, every text node part/.style={align=center}]
  \node[rect state,initial above]   (init)              {$\{(\x_{init},(0,0)),(\x_{init},(1,0)),(\x_{init},(2,0)),(\x_{init},(3,0))\}$};
  \node[rect state,accepting]   (0) [below=of init] {$\{(0,(0,0)),(1,(0,0)),(0,(1,0)),(0,(2,0)),(0,(3,0))\}$};
  \node[rect state,accepting]   (1) [below=of 0] {$\{(3,(0,0)),(4,(0,0)),(3,(1,0)),(2,(0,0)),(2,(1,0)),(3,(2,0)),(3,(3,0))\}$};
  \node[rect state]   (2) [below=of 1] {$\{(2,(0,0)),(2,(1,0)),(2,(2,0))\}$};
  \node[rect state,accepting]   (3) [below=of 2] {$\{(2,(0,0)),(2,(1,0)),(2,(2,0)),(2,(3,0))\}$};
  \path[->] (init) edge node [left] {$\sigma_{init}$} (0)
            (0) edge node [left] {$\sigma_o$} (1)
            (1) edge node [left] {$\sigma_o$} (2)
            (2) edge node [left] {$\sigma_o$} (3)
            (3) edge [loop below] node [below] {$\sigma_o$} ();
\end{tikzpicture}
\caption{The product(top) of $\G$ from Figure \ref{fig:transform_ex} with the nonsecret specification $\H_{NS,1}^{joint}(2)$ and the corresponding secret observer $\G_{SO}$ (bottom).}
\label{fig:joint_ex}
\end{figure}

For example we depict $\H_{NS,1}^{joint}(2)$ in  Figure \ref{fig:joint}.
Recall this automaton is constructed by concatenating $\H_*$ and three copies of $\H_{epoch,1}$.
We apply the secret observer method to the automaton $\A$ from Figure \ref{fig:transform_ex} using its label transform $\G = \autotrans(\A,\ell)$ also depicted in \ref{fig:transform_ex}.
The construction of $\G_{SO} = \det(\obs(\G \times \H_{NS,1}^{joint}(2)$ is depicted in Figure \ref{fig:joint_ex}.
We see that the string $\sigma_{init}\sigma_o \sigma_o$ is not marked in $\G_{SO}$.
Hence by the secret observer method, $\A$ is not jointly 2-step opaque with type 1 secrets.
Upon observing $\sigma_o \sigma_o$ we can deduce that $\A$ traversed the states $0,1,2,2$ or $0,3,4,2$ which both pass through secret states.

\subsection{Verification of separate $K$-step opacity}
\label{sec:verify_separate} 
Verification of separate $K$-step opacity is less straightforward than joint opacity.
Using the definition of separate opacity, we could do this by verifying the total opacity of $(\R,\R_{NS,j}(k))$ for each $k \leq K$ using the language-based methods.
Alternatively, we can combine these into a single test as in the joint case and avoid determinizing multiple automata by using two different approaches taking advantage of the structure of the problem.

By construction, $\H_{NS,j}(k)$ is embedded within $\H_{NS,j}(K)$ as a subautomaton for $k \leq K$.
So we can use $\H_{NS,j}(K)$ to specify the nonsecret runs $\R_{NS,j}(k)$ for $k \leq K$ for separate $K$-step opacity.
As in Remark \ref{rem:naming_conv}, we can write $\H_{NS,j}(k)=\bigcupdot_{i=0}^{k+1} \H^i$ where $\H^0 = \H_*, \ \H^1 = \H_{epoch,NS,j}$, and $\H^i = \H_{epoch}$ for $i \geq 2$.
Recall using the convention of Remark \ref{rem:naming_conv}, the marked states of $\H_{NS,j}(k)$ are simply the marked states of $H^k$ denoted by $\Q_{NS,m}^{k+1}$ embedded into $\H_{NS,j}(k)$ as $\Q_{NS,m}^{k+1} \times \{k+1\}$.
Hence it holds that $\lang_{\Q_{NS,m}^{k+1} \times \{k+1\}}(\H_{NS,j}(K)) = \lang_m(\H_{NS,j}(k))$.
 This yields the following approach.
\begin{approach}[Separate $K$-step opacity verification using secret observer]
\label{app:sep_ver_so}
Given $\A$, $\ell$, $\Sigma_o$, and $K < \infty$, construct  the label-transform $\G = \autotrans(\A, \ell)$, the nonsecret specification automaton $\H_{NS,j}(K)$, and the static mask $\obs$ induced by $\Sigma_o$.
Recall that $A$ is separate $K$-step opacity with type $j$ secrets if the $k$-delayed behavior with type $j$ secrets is opaque for each $k \leq K$.
We can verify this by applying the secret observer method  for each $k \leq K$  to $\G$, $\H_{NS,j}(K)$, and $\obs$ where we redefine the marked states of $\H_{NS,j}(K)$ to be $\Q_{NS,m}^{k+1} \times \{k+1\}$.
Each of these tests involves analyzing the states of the same automaton $\G_{SO} = \det(\obs(\G \times \H_{NS,j}(K)))$ under different notions of state markings.
As a result, we must only determinize a single automaton to apply this approach.
\hfill $\lozenge$
\end{approach}

However, the idea of this approach is not applicable to the reverse comparison method as this would require considering multiple sets of initial states.
Alternatively, we can avoid multiple determinizations by utilizing the fact that the intruder's knowledge of the system's behavior only increases as they make more observations.
Informally, if the intruder deduces a secret happened within the last $K-1$ observations, after making another observation they can still deduce a secret happened within the last $K$ observations.
So if the intruder can always make more observations, it suffices to consider secrets that occurred exactly $K$ observations ago for the purposes of verification.
This is similar to the results of Proposition 2 in \cite{sabooriNotionsSecurityOpacity2007}.
We will show under some conditions that it suffices to verify total opacity of $(\R,\R_{NS,j}(K))$ to verify separate $K$-step opacity with type $j$ secrets.
Here we say that $\A$ is \textbf{observation extendable} with respect to $\obs$ if for every $r \in \R = \ioseq(\A, \ell)$, there exists $r_{suf} \in \E_{uo}^* \E_o$ so that $(r\cdot r_{suf}) \in \R$, where $\E{uo},\E_o$ are defined as in equation \eqref{eq:observable_pair}.
With this we claim the following result.

\begin{theorem1}
\label{thm:obs_extend}
If $\A$ is observation extendable, then $\A$ is separate $K$-step opaque with type $j$ secrets if and only if $(\R,\R_{NS,j}(K))$ is totally opaque to $\obs$.
\end{theorem1}
\begin{proof}
Suppose that $\A$ is separately $K$-step opaque with type $j$ secrets.
Let $r \in \R$.
By the separate opacity of $\A$, there exists a run $r' \in \R_{NS,j}(K) = \R_{NS}$ with $\obs(r) = \obs(r')$.
Hence $(\R_S,\R_{NS,j}(K))$ is totally opaque to $\obs$.

Conversely, suppose that $(\R,\R_{NS,j}(K))$ is totally opaque to $\obs$.
Then let $r \in \R$ and $k \in \{0, \cdots, K\}$.
As $\R$ is observation extendable, there exists an extended run $r_{ext}=r \cdot r_{suf}$ so that $r_{ext} \in \R$ and $|\obs(r_{suf})| = K-k$.
By hypothesis, there exists a run $r'_{ext} \in \R_{NS} = \R_{NS,j}(K)$ with $\obs(r'_{ext}) = \obs(r_{ext})$.
By defining $r'_{suf}$ to be the last $K-k$ observation epochs of $r'_{ext}$, we can write $r_{ext}' = r' \cdot r_{suf}'$ with $|\obs(r_{suf}')| = K-k$.
Then we see that $r' \in \R_{NS,j}(k)$ and $\obs(r') = \obs(r)$.
Hence $\A$ is separately $K$-step opaque with type $j$ secrets.
\qed \end{proof}

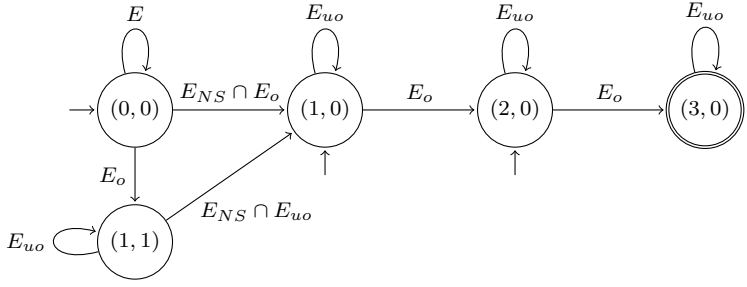
\begin{figure}
\centering
\begin{tikzpicture}[initial text=,shorten >=1pt,node distance=2.5cm,on grid, scale=1, transform shape, every text node part/.style={align=center}]
  \node[state,initial]   (init)                {$(0,0)$};
  \node[state,initial below]   (0NS)  [right=of init]    {$(1,0)$};
  \node[state]   (0S)  [below= 1.75cm of init]    {$(1,1)$};
  \node[state,initial below]   (1)  [right=of 0NS]    {$(2,0)$};
  \node[state,accepting]   (2)  [right=of 1]    {$(3,0)$};
  \path[->] (init) edge [loop above] node [above] {$\E$} ()
  				   edge node [above] {$\E_{NS} \cap \E_o$} (0NS)
  				   edge node [left] {$\E_o$} (0S)
  			(0NS)	   edge node [above] {$ \E_o$} (1)
  				   edge [loop above] node [above] {$\E_{uo}$} ()
  			(0S)   edge node [below right =0.25cm and -0.5cm] {$\E_{NS} \cap \E_{uo}$} (0NS)
  				   edge [loop left] node [left] {$\E_{uo}$} ()
  			(1)   edge [loop above] node [above] {$\E_{uo}$} ()
 			      edge node [above] {$\E_o$} (2)
  			(2) edge [loop above] node {$\E_{uo}$} ();
  \end{tikzpicture}
\caption{The nonsecret specification automaton $\H_{NS,2}(2)$ for separate 2-step opacity with type 2 secrets.}
\label{fig:separate}
\end{figure}

So when the system is observation extendable, we can verify separate $K$-step opacity in the following way.
\begin{approach}[Separate $K$-step opacity verification for observation extendable systems]
\label{app:sep_ver_ext}
Given $\A$, $\ell$, $\Sigma_o$, and $K < \infty$ where $\A$ is observation extendable with respect to the static mask $\obs$ induced by $\Sigma_o$, construct the label-transform $\G = \autotrans(\A, \ell)$ and the nonsecret specification automaton $\H_{NS,j}(K)$
We can verify the separate $K$-step opacity with type $j$ secrets of $\A$ by applying any of the language-based approaches to $\G$, $\H_{NS,j}(K)$, and $\obs$.
\hfill $\lozenge$
\end{approach}

\begin{remark1}
\label{rem:obs_ext}
While it may not be the case that $\A$ is observation extendable (for example if $\A$ is deadlocked), we can always modify $\A$ to be observation extendable while preserving $K$-step opacity.
To do this we define a new automaton $\A_{ext}$ by adding an artificial observable event $\sigma_{ext}$ as a self-loop for every state in $\A$.
Then one can show that $\A_{ext}$ will be separately $K$-step opaque if and only if $\A$ is.
Then by construction $\R_{ext}$ will be observation extendable, and so we can apply Approach \ref{app:sep_ver_ext} to $\A_{ext}$.
\hfill $\lozenge$
\end{remark1}

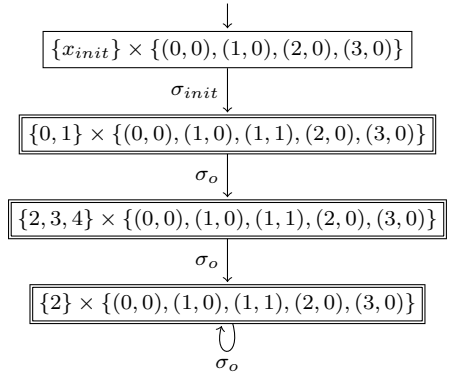
\begin{figure}
\centering

\begin{tikzpicture}[initial text=,shorten >=1pt,node distance=1.125cm,on grid, scale=1, transform shape, every text node part/.style={align=center}]
  \node[rect state,initial above]   (init)              {$\{\x_{init}\} \times \{(0,0),(1,0),(2,0),(3,0)\}$};
  \node[rect state,accepting]   (0) [below=of init] {$\{0,1\} \times \{(0,0),(1,0),(1,1),(2,0),(3,0)\}$};
  \node[rect state,accepting]   (1) [below=of 0] {$\{2,3,4\} \times \{(0,0),(1,0),(1,1),(2,0),(3,0)\}$};
  \node[rect state, accepting]   (2) [below=of 1] {$\{2\} \times \{(0,0),(1,0),(1,1),(2,0),(3,0)\}$};
  \path[->] (init) edge node [left] {$\sigma_{init}$} (0)
            (0) edge node [left] {$\sigma_o$} (1)
            (1) edge node [left] {$\sigma_o$} (2)
            (2) edge [loop below] node [below] {$\sigma_o$} ();
\end{tikzpicture}
\caption{The secret observer $\G_{SO}$ constructed for the automaton $\A$ from Figure \ref{fig:transform_ex} with the nonsecret specification $\H_{NS,2}(2)$.}
\label{fig:separate_ex}
\end{figure}

Using Approach \ref{app:sep_ver_ext} we can verify separate $K$-step opacity using the reverse comparison or secret observer method.
For example consider the system $\A$ from Figure \ref{fig:transform_ex} which is observation extendable and the nonsecret specification automaton $\H_{NS,2}(2)$ which is depicted in Figure \ref{fig:separate}.
The resulting secret observer $\G_{SO} = \det(\obs(\G \times \H_{NS,2}(2)))$ for $\G = \autotrans(\A,\ell)$ is depicted in Figure \ref{fig:separate_ex}.
As every state except the initial state is marked, we see that $\A$ is separately 2-step opaque with type 2 secrets.

\section{Complexity of K-step opacity verification}
\label{sec:complexity_k_step}
In this section, we analyze the complexity of the proposed methods for verifying $K$-step opacity for finite $K$ for an automaton $\A$ with labeling map $\ell$.
These methods use the transformed automaton $\G = \autotrans(\A,\ell)$.
First we analyze the secret observer using Approach \ref{app:joint_ver} for joint opacity and Approach \ref{app:sep_ver_so} for separate opacity.
Then we analyze the reverse language comparison using Approach \ref{app:sep_ver_ext}.
Finally, we compare the secret observer methods to existing verifiers for $K$-step opacity known as the $K$-delayed state and trajectory estimators \cite{sabooriVerificationKstepOpacity2009a, falconeRuntimeEnforcementKstep2013}.
For separate $K$-step we also compare with the two-way observer method \cite{yinNewApproachVerification2017}.
These results are summarized in Table \ref{tab:finite_res_sep} and Table \ref{tab:finite_res_joint}.

\subsection{Secret observer complexity}
\label{sec:complexity_so}
Recall that applying the secret observer method in Approach \ref{app:joint_ver}, \ref{app:sep_ver_so}, or \ref{app:sep_ver_ext} to verify $K$-step opacity involves constructing the automaton $\G_{SO} = \det(\obs(\G \times \H_{NS}))$ for an appropriate choice of $\H_{NS}$. 
We will bound the number of reachable states in this automaton to bound state complexity of these verification approaches.
A naive upper bound for the number of states in the power set construction for determinization of an automaton with $n$ states is simply $2^{n}$.
Using the known structure of $\H_{NS}$, we can obtain a tighter bound for determinizing the automaton $\obs(\G \times \H_{NS})$.
To do this, we will analyze which states of $\H_{NS}$ can be reached by runs that reach a fixed state of $\G$ in the following observation.

\begin{observation}
\label{obs:bound}
Consider two automata $\G = (\Q_{\G},\E,\f_{\G},\Q_{\G,0},\Q_{\G,m})$ and $\H_{NS} = (\Q_H,\E,\f_H,\Q_{\H,0},\Q_{\H,m})$ with a static mask $\obs: \E^* \rightarrow \Gamma^*$.
For convenience for $s \in E^*$ let $\f_{\G}(s) = \bigcup_{\q_{\G} \in \Q_{\G}} \f_{\G}(\q_{\G},s)$ and $\f_{\H}(s) = \bigcup_{\q_{\H} \in \Q_{\H}} \f_{\H}(\q_{\H},s)$.
Suppose we are given sets $F \subseteq 2^{\Q_H}$ and $C \subseteq \Gamma^*$ such that $F$ is closed under union, $\emptyset \in F$, and for all $s \in \lang(\G\times H)$ such that $\obs(s) \in C$ it holds that $\f_H(s) \in F$.
Then for every $\gamma \in C$ we can define the function $w_{\gamma} : \Q_{\G} \rightarrow F$ by \begin{equation}
w_{\gamma}(\q_{\G}) = \bigcup_{\substack{s \in \obs^{-1}(\gamma) \\ s.t. \ \q_{\G} \in \f_{\G}(s)}} \f_H(s)
\end{equation}
Then denote the automaton $\obs(\G \times \H_{NS})$ as \begin{equation}
\obs(\G \times \H_{NS}) = Q_{\obs(\G \times H)}, \Gamma \cup \{\epsilon\}, \f_{\obs(\G \times H)}, \Q_{\obs(\G \times H),0}, \Q_{\obs(\G \times H),m}) \, .
\end{equation}
For $\gamma \in C$ it holds that \begin{equation}
\f_{\obs(\G \times H)}(\gamma) = \bigcup_{s \in \obs^{-1}(\gamma)} \f_{\G}(s) \times \f_H(s) = \bigcup_{\q_{\G} \in \Q_{\G}} \left( \{\q_{\G}\} \times w_{\gamma}(\q_{\G}) \right) \, .
\end{equation}
Hence the number of states in $\det(\obs(\G \times \H_{NS}))$ reached by a string in $C$ is bounded by the number of functions from $\Q_{\G}$ to $F$, of which there are $|F|^{|\Q_{\G}|}$.
\hfill $\lozenge$
\end{observation}

We can apply this observation to bound the complexity of the secret observer method.
As $\G_{SO} = \det(\obs(\G \times \H_{NS}))$ is deterministic, it has a single initial state reached by $\epsilon$.
So all states of $\G_{SO}$ other than the initial state are reached by $C = \Gamma^+$.
Then to apply Observation \ref{obs:bound}, we must determine a set $F \supseteq \{\f_{\H_{NS}}(s) \mid s \in C\}$ which is also closed under union and contains the empty set.
We claim in verifying joint $K$-step opacity that
\begin{itemize}
\item for $\H_{NS} = \H_{NS,1}^{joint}(K)$, we can choose $|F| = K+3$ with \begin{equation}
F = \{\emptyset\} \cup \{ \{0,\cdots,k\} \times \{0\} \}_{k=0}^{K+1}\, ,
\end{equation}
\item for $\H_{NS} = \H_{NS,2}^{joint}(K)$,  we can choose $|F| = 2(K+1)+1$ with 
\begin{multline}
F =  \{\emptyset\} \cup 
\{\{(0,0),(k+1,1)\} \cup (\{1,\cdots,k\} \times \{0,1\})\}_{k=0}^{K} \cup \\
\{\{(0,0)\} \cup (\{1,\cdots,k+1\} \times \{0,1\})\}_{k=0}^K \, .
\end{multline}
\end{itemize}
If we denote the number of states of the original automaton $\A$ as $n = |X|$, then the number of states of $\G = \autotrans(\A,\ell)$ is $n+1$, including the artificial initial state.
Observation \ref{obs:bound} then shows the number of states of $\G_{SO}$ other than the initial state is bounded by $|F|^{n}$.
These bounds are given by $(K+3)^n$ for $\H_{NS,1}^{joint}(K)$ and $(2K+3)^n$ for $\H_{NS,2}^{joint}(K)$.
For separate opacity, we use the naive power set bounds of $2^{n(K+2)}$ for $\H_{NS,1}(K)$ and $2^{n(K+3)}$ for $\H_{NS,2}^{joint}(K)$.
These bounds are summarized in Table \ref{tab:finite_res_sep} and Table \ref{tab:finite_res_joint}.

\subsection{Reverse comparison complexity}
\label{sec:complexity_rc}
We can use the same approach to analyze the reverse comparison method as in Approach \ref{app:joint_ver} and Approach \ref{app:sep_ver_ext} to verify $K$-step opacity. 
These approaches require constructing the automaton $\G_{RC} = \obs(\G)^R \times \det(\obs(\G \times \H_{NS})^R)$ for an appropriate choice of $\H_{NS}$.
By observing that $\obs(\G \times \H_{NS})^R = \obs(\G^R \times \H_{NS}^R)$, we can use Observation \ref{obs:bound}, to bound the number of reachable states of $\det(\obs(\G^R \times \H_{NS}^R)$.
For the nonsecret specification automata $\H_{NS}$ use for $K$-step opacity, the reachable sets of $\H_{NS}^R$ are simpler than $\H_{NS}$.
Consider a string $s \in (L_{epoch}^+)^R$ with $k = \max(0,K+1-|\obs(s)|)$.
Using the notation from Remark \ref{rem:naming_conv}, we can see that $\H_{NS}$ must reach a state corresponding to $\H_{NS}^k$.
Consider the set $C_k = \Gamma^{K+1-k}$ with $1 \leq k \leq K$ and $C_0 = \Gamma^{K+1} \Gamma^*$.
Then we determine a set $F_k \supset \{\delta_{\H_{NS}^R}(s) \mid s \in C_k\}$ that is closed under union and contains the empty set.
We claim that \begin{itemize}
\item for $\H_{NS} = \H_{NS,1}^{joint}$ or $\H_{NS} = \H_{NS,1}$ we can choose $|F_k| = 2$ with \begin{equation}
F_k = \{\{(k,0)\}, \emptyset \} \, .
\end{equation}
\item for $\H_{NS} = \H_{NS,2}^{joint}$ or $\H_{NS} = \H_{NS,2}$ we can choose $|F_k| = 3$ with \begin{equation}
F_k = \{\{(k,0)\}, \{(k,0),(k,1)\}, \emptyset \}
\end{equation}
\end{itemize}
So by Observation \ref{obs:bound} for $C_k$, the number of states of $\det(\obs(\G^R \times \H_{NS}^R))$ reached by a string $\gamma \in \Gamma^+$ with $k = \max(0,K+1-|\gamma|)$ is bounded by $|F_k|^{n+1}$ where $n=|X|$ is the number of states in the original automaton $\A$.
Hence the number of states of $\det(\obs(\G^R \times \H_{NS}^R))$ is  $O((K+1)2^n)$ for type 1 secrets and $O((K+1)3^n)$ for type 2 secrets.
So then the number of states of $\G_{RC} = \obs(\G^R) \times \obs(\det(\G \times \H_{NS}))^c$ is $O(n(K+1)2^n)$ for $\H_{NS} = \H_{NS,1}(K),\H_{NS,1}^{joint}(K)$ and $O(n(K+1)3^n)$ for $\H_{NS} = \H_{NS,2}(K),\H_{NS,2}^{joint}(K)$.
These bounds are depicted in Table \ref{tab:finite_res_sep} and Table \ref{tab:finite_res_joint}.

To demonstrate the advantage of the reverse language comparison, consider the following family of automata.
Define for $n > 1$, $\A(n) = (\X_n,\Sigma_n,\delta_n,\X_n \setminus \{0\}, \X_n,A,\ell)$ where $\X_n = \{0,\cdots,n-1\}$, $\Sigma_n = \{\sigma_0,\cdots, \sigma_{n-1}\}$,  $\delta_n(i,\sigma_j) = (i+j) \bmod n$, $A=\{S,NS\}$, and $\ell_n(0) = S$ and $\ell_n(i) = NS$ for $i \neq 0$.
We define all events to be observable $\Sigma_o = \Sigma_n$.
After constructing $\G(n) = \autotrans(\A(n),\ell_n)$ for various $n$, we compute the number of states in the secret observer automaton $\G_{SO}(n) = \det(\obs(\G(n) \times \H_{NS}))^c$ and in the reverse automaton $\G_{RC}(n)  = \obs(\G(n)^R) \times \det(\obs(\G(n)^R \times \H_{NS}^R))^c$ for $\H_{NS} = \H_{NS,1}^{joint}(K)$ and $\H_{NS} = \H_{NS,2}(K)$ across various values of $K$.
These results are depicted in Figure \ref{fig:forward_reverse_comp}.
The number of states in the forward automata increases roughly exponentially with $K$ while the number of states in the reverse automata increases linearly.

\begin{figure}
\centering
\begin{tabular}{|c|c|c|c|c|}
\hline
$K$ & Forward ($n=4$) & Reverse ($n=4$) & Forward ($n=6$) & Reverse ($n=6$) \\ \hline
0 & 5 & 6 & 7 & 8 \\ \hline
2 & 53 & 29 & 187 & 67 \\ \hline
4 & 293 & 45 & 3007 & 147 \\ \hline
8 & 2117 & 77 & 114487 & 275 \\ \hline
16 & 16517 & 141 & T/O & 531 \\ \hline
\end{tabular}

\vspace{0.25cm}
\begin{tabular}{|c|c|c|c|c|}
\hline
$K$ & Forward ($n=4$) & Reverse ($n=4$) & Forward ($n=6$) & Reverse ($n=6$) \\ \hline
0 & 5 & 6 & 7 & 8 \\ \hline
2 & 35 & 29 & 137 & 67 \\ \hline
4 & 137 & 45 & 1547 & 147 \\ \hline
8 & 749 & 77 & 36047 & 275 \\ \hline
16 & 4949 & 141 & 1071767 & 531 \\ \hline
\end{tabular}
\caption{The number of states in the forward secret observer automata $\G_{SO}(n)$ and reverse automata $\G_{RC}(n)$ constructed from $\G(n) = \autotrans(\A(n),\ell_n)$.
The bottom table uses $\H_{NS} = \H_{NS,1}^{joint}(K)$ and the top table uses $\H_{NS} = \H_{NS,2}(K)$.
Here T/O denotes a timeout where the automaton could not be constructed.}
\label{fig:forward_reverse_comp}
\end{figure}

\subsection{Comparison to $K$-delay State \& trajectory estimators}
\label{sec:k_step_comparison}
We can explicitly compare our secret observer method with some existing methods for verification of $K$-step opacity.
We consider weak and strong $K$-step opacity over a deterministic automaton $\A = (\X,\Sigma,\delta,\{\hat{x}_0\})$ with secret states $X_S$ defining a label map $\ell$..
The first proposed verification methods for weak and strong $K$-step opacity are called the $K$-delay state estimator and $K$-delay trajectory estimator.
These $K$-delay state estimator constructing an automaton that estimates the possible states sequences over the last $K$ observations from which one can deduce if weak opacity has been violated.
The $K$-delay trajectory estimator augments this structure with a sequence of binary variables representing whether or not a secret state was visited between the observations.
Their complexities are depicted in Table \ref{tab:finite_res_sep} and Table \ref{tab:finite_res_joint}.
We construct a map $g$ from states of these estimators into the states of our secret observer automaton $\G_{SO}$ for $\G = \autotrans(\A,\ell)$ for either weak or strong $K$-step opacity.
We show that if the state $\x_{est}$ is reached by the string $s$ in the $K$-delay estimator, then the state $g(\x_{est})$ is reached by $\sigma_{init} s$ in $\G_{SO}$.
As both of these automata are deterministic and generate the same languages, ignoring the initial event $\sigma_{init}$, the number of non-initial states of $\G_{SO}$ is no more than the number of states of the corresponding $K$-delay estimator.

First we consider verifying weak $K$-step opacity with the state-mapping-based $K$-delay estimator denoted by $\A_{K,obs} = (\X_{K,obs}, \Sigma_o,\delta_{K,obs}, \{\x_{K,obs,0}\},$ $\X_{K,obs})$ from \cite{sabooriVerificationKstepOpacity2009a}.
We will assume that the initial state of $\A$ has no outgoing unobservable events, but we can extend the following argument to a general $\A$.
In this case $\A_{K,obs}$ estimates all possible tuples $(\x_0,\cdots,\x_K) \in \X^{K+1}$ so that for $i \geq K$ it holds that $\x_i$ was visited between $K-i$ and $K-i-1$ observations ago, and for $i < K$ it holds that $\x_i = \hat{x}_0$.
In this case, the initial state singleton of the estimator is $\{\x_{K,obs,0}\} = \{\x_0\}^{K+1}$.
Note if $\ell(\x_0)=S$ then this system is trivially not opaque, so we will consider when $\ell(\x_0) = NS$.
To verify weak opacity with the secret observer method as in Approach \ref{app:sep_ver_so}, we construct the secret observer $\G_{SO} = \det(\obs(\G \times \H_{NS,2}(K)))$ which is denoted $\G_{SO} = (\Q_{SO},\Gamma, \f_{SO},\{q_{SO,0}\}, \Q_{SO,m})$,
where $\G = \autotrans(\A,\ell) = (Q,E,f,Q_0,Q_m)$ and $\H_{NS,2}(K) = (\Q_H,E,f_{\H},\Q_{\H,0},\Q_{\H,m})$.
Recall $\Q = \X \cup \{\x_{init}\}$.
We then have the following result.
\begin{theorem1}
\label{thm:weak_comparison}
Define the map $g_{weak}:2^{\X^{K+1}} \rightarrow 2^{\Q \times \Q_H}$ \begin{multline}
g_{weak}(S) = \{(\q,\q^H) \in \Q \times \Q_H \mid \exists (\x_0, \cdots, \x_K) \in S, \ \x_K = \q, \\ \q^H \in \{(0,0),(1,1)\} \cup \{(k,0) \mid \ell(\x_{K-k}) = NS\}\}\, .
\end{multline}
Using the convention of Remark \ref{rem:naming_conv}, we enumerate the states of $\H_{NS,2}(K)$ as $\Q_H = \{(0,0),(1,0),(1,1),(2,0),\cdots,(K+1,0)\}$.
Then for every $\gamma \in \lang(\A_{K,obs})$, it holds that \begin{equation}
g_{weak}(\delta_{K,obs}(\gamma,x_{K,obs,0})) = \f_{SO}(\q_{SO,0}, \sigma_{init} \cdot \gamma) \, .
\end{equation}
\end{theorem1}
\begin{proof}
Let $\gamma \in \lang(\A_{K,obs})$.
Consider a sequence $(\x_0,\cdots,\x_K) \in \delta_{K,obs}(\gamma,\x_{K,obs,0})$.
By definition of $\A_{K,obs}$, there must exist a string $s \in \lang(\A)$ with $P_{\Sigma_o}(s) = \gamma$ generating a state trajectory corresponding to this estimate.
Then there must exist a corresponding run of input-output pairs $r \in \R = \ioseq(\A,\ell)$  reaching the state $\x_K$ in $\G$ with  $\obs(r) = \sigma_{init} \cdot \gamma$.
Additionally, dividing $r$ into epochs as $r=r_{epoch,0}\cdots r_{epoch,M}$ with $r_{epoch,i} \in L_{epoch}$, it holds that $r_{epoch,(M-k)} \in L_{epoch,NS,2}$ if $\ell(\x_{K-k})$ for $k \in \{0,\cdots, \max(K,m)\}$.
So then $r$ can reach any state of $g_{weak}(\{\x_0,\cdots,\x_K\})$ in $\G \times \H_{NS}$.
Hence $g_{weak}(\delta_{K,obs}(\gamma,\x_{K,obs,0})) \subseteq  \f_{SO}(\q_{SO,0},\sigma_{init} \cdot \gamma)$.
Reversing this argument yields the converse. 
\qed \end{proof}

Next we consider verifying strong $K$-step opacity with the $K$-delayed trajectory estimator denoted  $\A_{K,obs}' = (\\X_{K,obs}',\Sigma_o,\delta_{K,obs}',\{\x_{K,obs,0}'\})$  from \cite{falconeRuntimeEnforcementKstep2013}.
For simplicity as in the weak opacity setting, we assume that the initial state of $\A$ has no outgoing unobservable events and that this initial state is nonsecret.
This automaton estimates the state tuple $(\x_0, \cdots, \x_K) \in \X^{K+1}$ as in the weak case, along with a binary tuple $(b_0,\cdots,b_{K-1}) \in \{0,1\}^{K}$ where $b_i$ represents whether or not the partial trajectory between $\x_i$ and $\x_{i+1}$ visited a nonsecret state.
To verify strong opacity of $\A$ with the secret observer method as in Approach \ref{app:joint_ver}, we construct the secret observer automaton $\G_{SO} = \det(\obs(\G \times \H_{NS,1}^{joint}(K)))$ which is denoted $\G_{SO} = (\Q_{SO},\Gamma, \f_{SO},\{q_{SO,0}\}, \Q_{SO,m})$,
where $\G = \autotrans(\A,\ell) = (Q,E,f,Q_0,Q_m)$ and $\H_{NS,1}^{joint}(K) = (\Q_H,E,f_{\H},\Q_{\H,0},\Q_{\H,m})$.
Recall $\Q = \X \cup \{\x_{init}\}$.
We then have the following result.
\begin{theorem1}
\label{thm:strong_comparison}
Define the map $g_{strong}:2^{\X^{K+1} \times \{0,1\}^K} \rightarrow 2^{\Q \times \Q^H}$ \begin{multline}
g_{strong}(S) = \{(\x_K,\q^H) \in \Q \times \Q^H \mid \exists (\x_0,\cdots,\x_K,b_0,\cdots,b_{K-1}) \in S, \\ \x^H \in \{0\} \cup \{k+1 \mid \ell(\x_K)=NS \ \wedge \ \forall i \geq K-k \ b_{i}=0\}\}\, .
\end{multline}
Using the convention of Remark \ref{rem:naming_conv}, we enumerate the states of $\H_{NS,1}^{joint}(K)$ as $\Q_H = \{(0,0),(1,0),\cdots,(K+1,0)\}$, and consider the secret observer $\G_{SO}= \det(\obs(\G \times \H_{NS})$.
Then for every $\gamma \in \lang(\A_{K,obs})$, it holds that \begin{equation}
g_{strong}(\delta_{K,obs}(\gamma,\x_{K,obs,0})) = \f_{SO}(\q_{SO,0}, \sigma_{init} \cdot \gamma) \, .
\end{equation}
\end{theorem1}
\begin{proof}
The proof is similar to the weak case
\qed \end{proof}

These results show that the number of states in the relevant $K$-delay state or trajectory estimators is at least the number of non-initial states in the corresponding secret observer.
To demonstrate this, we construct a family of automata $\A(i)$ where the secret observer method has significantly reduced complexity compared to the delayed state/trajectory method for verification of strong/weak $K$-step opacity.
For $i > 1$ define the deterministic automaton $\A(i) = (\X_i,\Sigma_i,\delta_i,\{2\})$ where $\X_i = \{1,\cdots,i\}$, $\Sigma_i = \{\sigma_1, \cdots, \sigma_i\}$, $\Sigma_o = \Sigma$, and the transition function defined by $\delta_i(j,\sigma_k) = k$.
Consider the labeling map $\ell_i:\X_i \rightarrow A$ where $A = \{S,NS\}$ defined by $\ell_i(1) = S$ and $\ell_i(j) = NS$ for $j \neq 1$.
Note that $\A(i)$ recognizes a run along every state sequence in $\{2\}\cdot(\X_i)^*$.
Hence we see the $K$-delayed state observer states correspond to $\bigcup_{k=0}^{K} \{2\} \times (\X_i)^{k}$, of which there are $\sum_{k=0}^{K} i^k = \frac{1-i^{K+1}}{1-i} = O(i^{K})$ states.
Let $\G(i) = \autotrans(\A(i), \ell_i)$.
The secret observer $\G_{SO}(i) = \det(\obs(\G(i) \times \H_{NS,2}(K)))$ estimates the current state and the secrecy of the past $K+1$ epochs. 
We can verify that the number of states in $\G_{SO}(i)$ is $O(i 2^K)$.

So we see that the secret observer method can be significantly less complex than the delayed state estimator for verification of weak $K$-step opacity.
A similar result holds for strong $K$-step opacity.

\begin{table}
\centering
\begin{tabular}{|c|c|}
\hline 
Separate Type 2 (Weak) & \\ \hline
Algorithm & State Complexity \\ \hline
Secret Observer & $O(2^{n(K+3)})$ \\ \hline
Reverse Comparison & $O(n(K+1)3^n)$ \\ \hline
State Estimator \cite{sabooriVerificationKstepOpacity2009a} & $O((|\Sigma_o|+1)^K 2^n)$ \\ \hline
Two-way Observer \cite{yinNewApproachVerification2017} & $O(\min(2^n, |\Sigma_o|^K) 2^n)$ \\ \hline
\end{tabular}
\caption{State complexities of verification methods for separate $K$-step opacity with type 2 secrets (weak $K$-step opacity) of an automaton with $n$ states.
Theorem \ref{thm:weak_comparison} implies that the secret observer method has state complexity no worse than the $K$-delay state estimator.}
\label{tab:finite_res_sep}
\end{table}

\begin{table}
\centering
\begin{tabular}{|c|c|}
\hline 
Joint Type 1 (Strong) & \\ \hline
Algorithm & State Complexity \\ \hline
Secret Observer & $O((K+3)^n)$ \\ \hline
Reverse Comparison & $O(K 2^n)$ \\ \hline
Trajectory Estimator \cite{falconeRuntimeEnforcementKstep2013} & $O((|\Sigma_o|+1)^K 2^n)$ \\ \hline
\end{tabular}
\caption{State complexities of verification methods for joint $K$-step opacity with type 1 secrets (strong $K$-step opacity) of an automaton with $n$ states.
Theorem \ref{thm:strong_comparison} implies that the secret observer method has state complexity no worse than the $K$-delay trajectory estimator.}
\label{tab:finite_res_joint}
\end{table}

\section{Infinite step opacity}
\label{sec:inf_step}
Now we consider $K$-step opacity for $K=\infty$, also called infinite step opacity.
The results of Theorem \ref{thm:k_step_equivalence} can be extended to the infinite step case.
In particular our notion of separate infinite-step opacity with type 2 secrets corresponds to the existing notion of infinite step opacity as in \cite{SABOORI200946, yinNewApproachVerification2017}.
We will discuss how the previous verification methods for finite $K$ can be adapted to this infinite case.

Recall our definition of infinite step opacity involves an infinite number of nonsecret language specifications, i.e. the $k$-delayed nonsecret behavior $L_{NS,j}(k)$ for $k \in \mathbb{N}$ as defined in \eqref{eq:k_step_behavior}.
Recall in the finite case we were able to reduce the multiple language comparison checks into a single check for verifying separate opacity.
In Approach \ref{app:sep_ver_so}, we constructed one automaton that encompassed all of the nonsecret behavior, but this automaton would necessarily be infinite for $K=\infty$.
In Approach \ref{app:sep_ver_ext}, under the condition of observation extendability we showed it suffices to consider secret behavior occurring exactly $K$ epochs ago, but there is no clear analog for this for $K=\infty$.
Hence it appears that we cannot directly use our methods for verification of separate infinite step opacity.
However we can use a result of \cite{yinNewApproachVerification2017} that states that infinite step opacity (separate opacity with type 2 secrets) is equivalent to $K$-step opacity for $K = 2^n$ where $n$ denotes the number of states of the automaton in question.
With this observation, we can verify separate infinite step opacity with type 2 secrets by verifying separate $2^n$-step opacity.
Alternatively, the two-way observer could be used to directly verify separate infinite step opacity \cite{yinNewApproachVerification2017}.

We can more effectively apply our methods to joint infinite step opacity as this involves only one language comparison by definition.
Note that we can define \begin{equation}
L_{NS,j}^{joint}(\infty) = \bigcap_{i=0}^\infty L_{NS,j}(i) = L_{epoch,NS,j}^+
\end{equation}

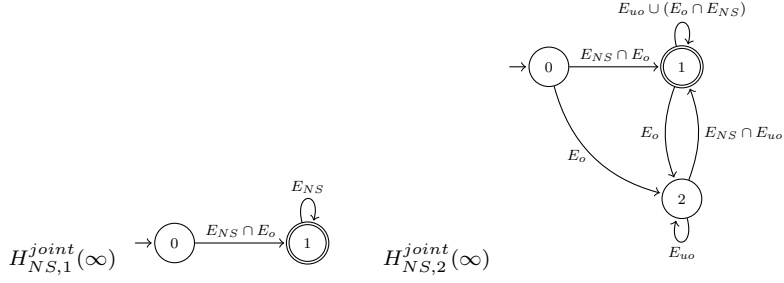
\begin{figure}
\centering
$\H_{NS,1}^{joint}(\infty)$ \begin{tikzpicture}[initial text=,shorten >=1pt,node distance=2.5cm,on grid, scale=0.7, transform shape, every text node part/.style={align=center}]
  \node[state,initial]   (0)                {$0$};
  \node[state, accepting]   (1) [right=of init]  {$1$};
  \path[->] (0) edge node [above] {$\E_{NS} \cap \E_o$} (1)
  			(1) edge [loop above] node [above] {$\E_{NS}$} ();
\end{tikzpicture}
\qquad $\H_{NS,2}^{joint}(\infty)$ \begin{tikzpicture}[initial text=,shorten >=1pt,node distance=2.5cm,on grid, scale=0.7, transform shape, every text node part/.style={align=center}]
  \node[state,initial]   (init)                {$0$};
  \node[state, accepting]   (0) [right=of init]  {$1$};
  \node[state]   (1) [below=of 0]  {$2$};
  \path[->] (init) edge node [above] {$\E_{NS} \cap \E_o$} (0)
                   edge [bend right=30] node [left] {$\E_o$} (1)
  			(0) edge [loop above] node [above] {$\E_{uo} \cup (\E_o \cap \E_{NS})$} ()
  			    edge [bend right=20] node [left] {$\E_o$} (1)
  			(1) edge [loop below] node [below] {$\E_{uo}$} ()
  			    edge [bend right=20] node [right] {$\E_{NS} \cap \E_{uo}$} (0);
\end{tikzpicture}
\caption{The nonsecret specification automata for joint infinite step opacity.}
\label{fig:spec_auto_joint_inf}
\end{figure}

As in the finite case, we can construct an automaton to specify this nonsecret behavior.
Consider the automata depicted in Figure \ref{fig:spec_auto_joint_inf}.
Note that $\lang_m(H_{NS,j}^{joint}(\infty)) \neq \E^*$ so these cannot be used with the secret observer method as we have done before.
We can, however, analyze the complexity of the forward comparison method using Observation \ref{obs:bound}.
Consider the set $C = \Gamma^+$.
Then we determine a set $F \supset \{\delta_{\H_{NS}}(s) \mid s \in C_k\}$ that is closed under union and contains the empty set.
We claim that
\begin{itemize}
\item for $\H_{NS} = \H_{NS,1}^{joint}(\infty)$ we can choose $|F| = 2$ with \begin{equation}
F = \{\emptyset, \{1\} \} \, .
\end{equation}
\item for $\H_{NS} = \H_{NS,2}^{joint}(\infty)$ we can choose $|F| = 3$ with \begin{equation}
F = \{\emptyset, \{2\}, \{1,2\} \}
\end{equation}
\end{itemize}
So by Observation \ref{obs:bound} for $C$, the number of states of $\det(\obs(\G \times \H_{NS}))$ reached by a string $\gamma \in \Gamma^+$ is bounded by $|F|^n$ with $n=|\X|$ where $\G = \autotrans(\A, \ell)$.
Then the number of states in the automaton $G_{FC} = \obs(\G) \times \det(\obs(\G \times \H_{NS}))^c$ other than the initial state is $O(n 2^n)$ for type 1 secrets and $O(n 3^n)$ for type 2 secrets.
To the best of our knowledge, verification of joint infinite step opacity has not been reported in the literature previously. 

\section{Numerical examples}
\label{sec:examples}
We evaluate the effectiveness of our verification methods for $K$-step opacity with numerical experiments.
We compare the time and space complexity of the proposed methods with existing methods for verifying the existing notions of strong and weak $K$-step opacity.
Recall these correspond to the notions of joint $K$-step opacity with type 1 secrets and separate $K$-step opacity with type 2 secrets, respectively.
It should be noted while the existing methods were originally described for deterministic automata, there is a natural extension to the nondeterministic automata considered here.
We compare the runtimes and number of states in the final verifier automata for an implementation of each method.
In order to show how these methods scale with the size of the original system and the value of $K$, we verify the opacity of systems represented by randomly generated automata with secret states.
We generate these automata in two ways.
We present the runtimes and number of states in the verification automata averaged over 100 systems for fixed system sizes up to 250 states .
These methods were implemented in the \textit{DESops} library \footnote{The library is available at \url{https://gitlab.eecs.umich.edu/M-DES-tools/desops/} .}.

\begin{figure}[t]
\centering
\includegraphics[scale=0.45]{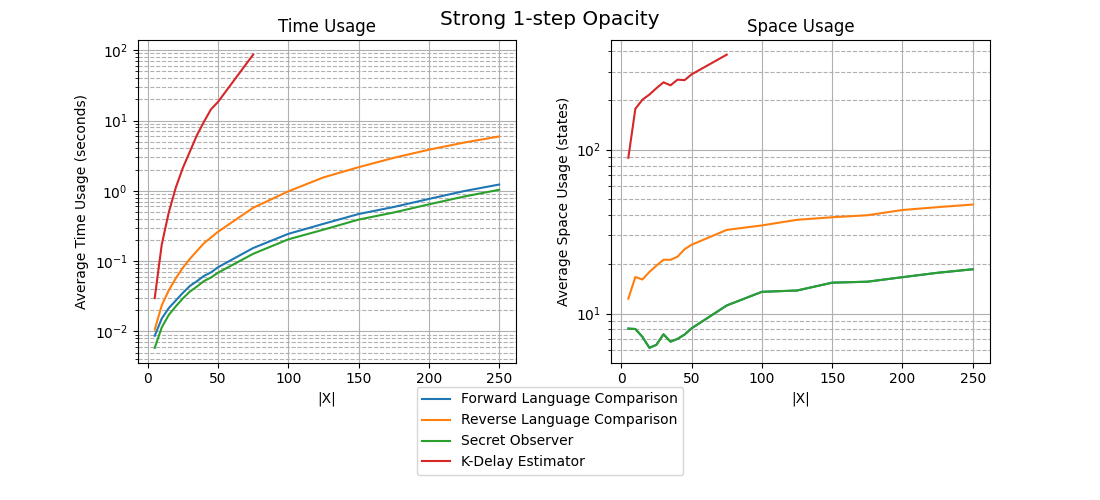}
\includegraphics[scale=0.45]{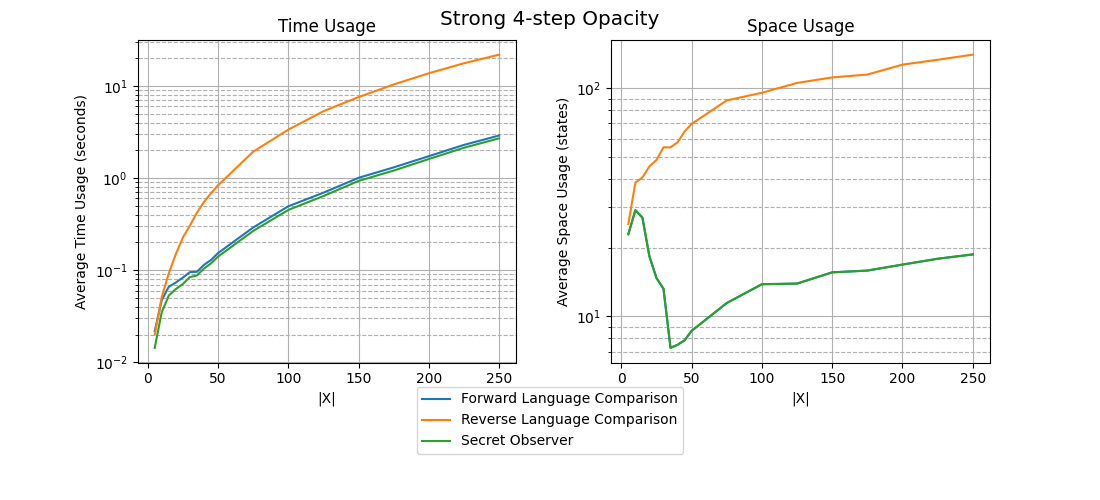}
\caption{Plots of average runtime (time usage) and the number of states in the verifier automata (space usage) versus the number of states in the random automata system model ($|\X|$) for several methods for verifying strong $K$-step opacity.}
\label{fig:strong_exp}
\end{figure}

\subsection{First random generation approach}

For the first experiment, we generate automata with a fixed number of states with a random number of outgoing transitions to random states.
There are 18 events total with 6 observable events.
All states are considered to be initial, and one state is labeled as secret.

\begin{figure}[t]
\centering
\includegraphics[scale=0.45]{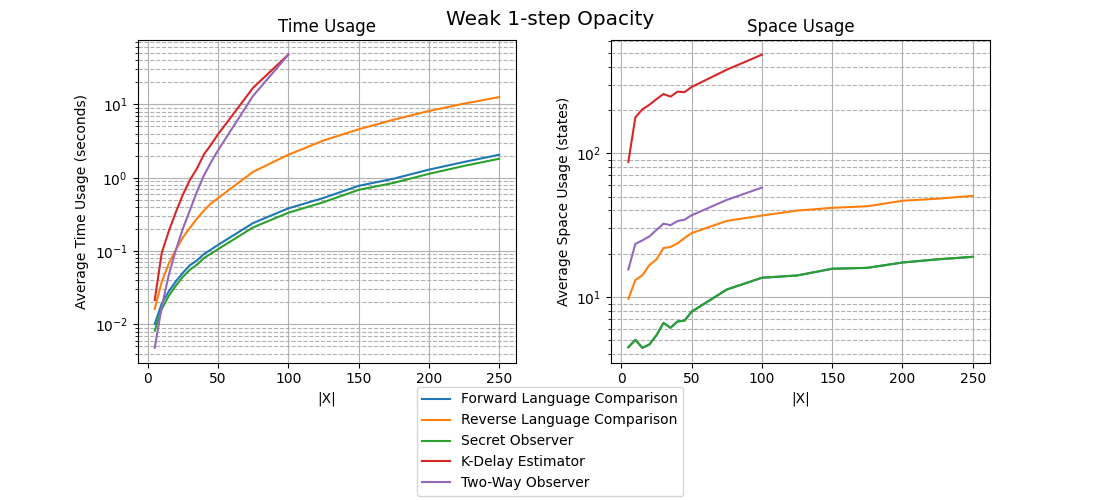}
\includegraphics[scale=0.45]{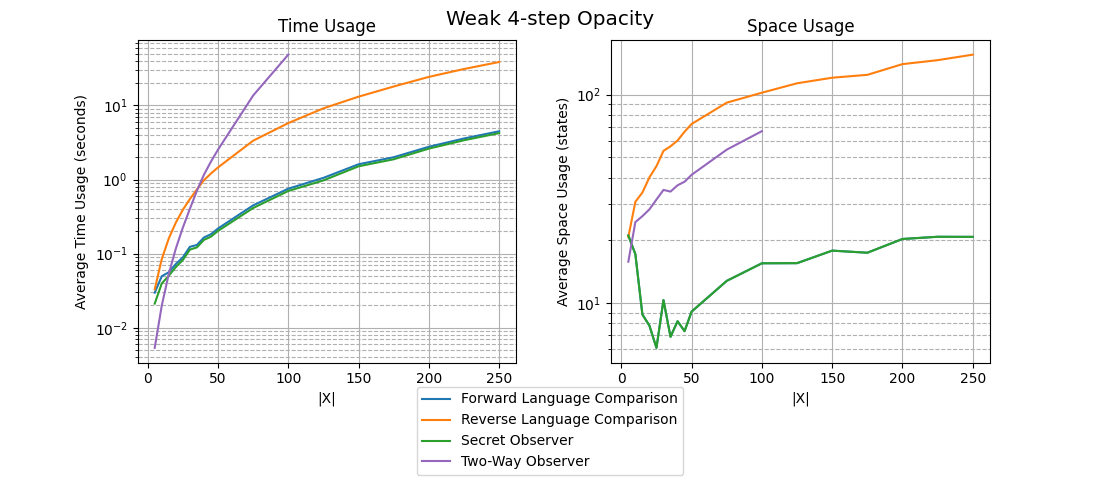}
\caption{Plots of average runtime (time usage) and the number of states in the verifier automata (space usage) versus the number of states in the random automata system model ($|\X|$) for several methods for verifying weak $K$-step opacity.}
\label{fig:weak_exp}
\end{figure}

For strong $K$-step opacity, we compare the proposed forward comparison, reverse comparison, and secret observer methods with the existing $K$-delay trajectory estimator.
We consider both $K=1$ and $K=4$.
The average results over the randomly generated automata for verifying strong $K$-step opacity are depicted in Figure \ref{fig:strong_exp}.
Due to the long runtime of the $K$-delay trajectory estimator ($>100s$), we do not evaluate this method for large automata in the $K=1$ case and remove it entirely in the $K=4$ case.
In these examples, the forward comparison method performed nearly identically to the secret observer method, which is why it does not appear in the space usage plots.
From these plots, we see that the proposed methods for verification perform significantly  faster than the existing method.
This affirms the results of Theorem \ref{thm:strong_comparison}, stating that the complexity of the secret observer method for verifying $K$-step opacity is less than that of the $K$-delay trajectory estimator.
It is also interesting to note that the secret observer method outperforms the reverse language comparison for the small values of $K$ investigated.
This indicates the linear scaling with $K$ in the complexity of this method is only significant for large values of $K$.

\begin{figure}[t]
\centering
\includegraphics[scale=0.45]{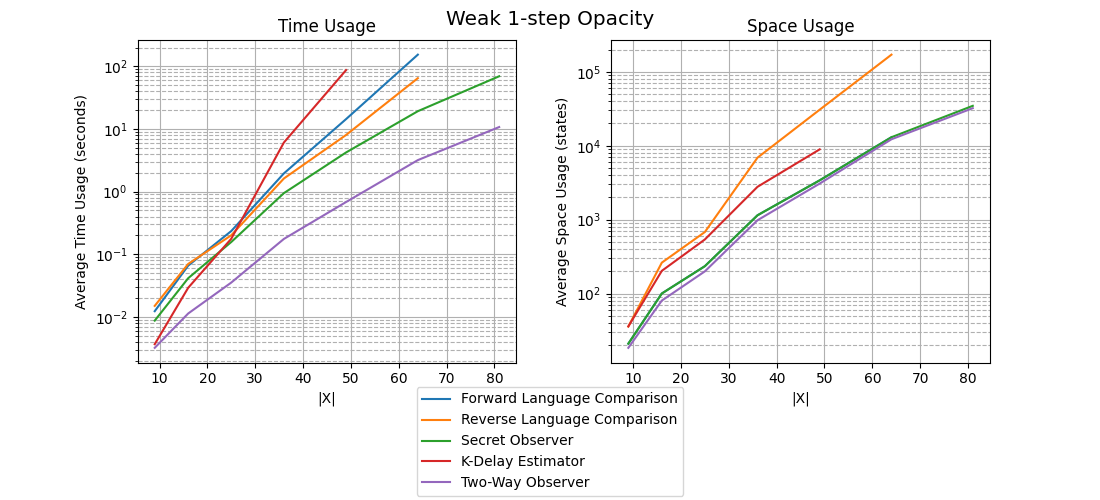}
\includegraphics[scale=0.45]{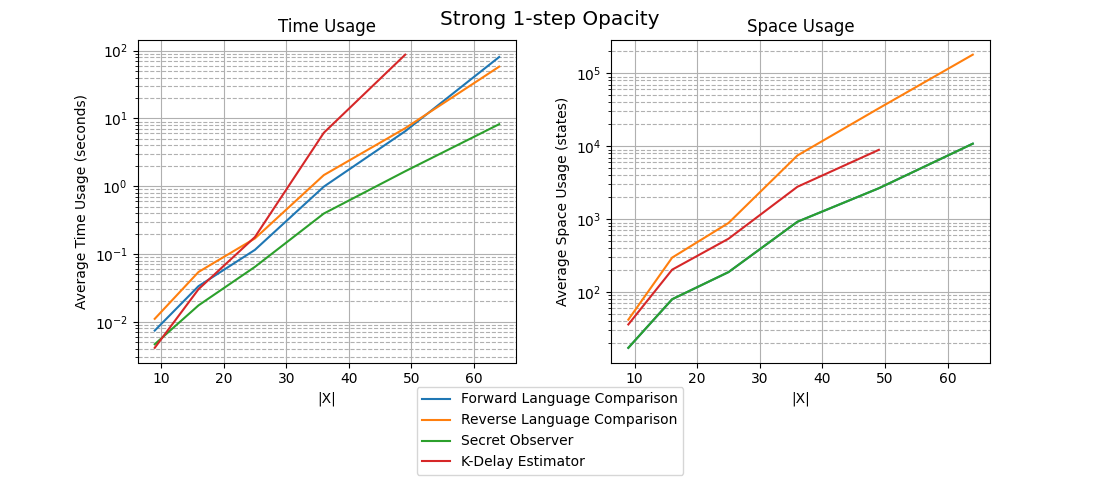}
\caption{Plots of average runtime (time usage) and the number of states in the verifier automata (space usage) versus the number of states in the system model ($|\X|$) for several methods for verifying strong and weak $1$-step opacity for the grid-based automata.}
\label{fig:grid_based}
\end{figure}

For weak $K$-step opacity, we compare the proposed forward comparison, reverse comparison, and secret observer methods with the existing $K$-delay state estimator and the two-way observer \cite{yinNewApproachVerification2017}.
For the secret observer method, Approach \ref{app:sep_ver_so} is used, while for the forward and reverse comparison methods, Approach \ref{app:sep_ver_ext} is used.
As in the strong case, we consider both $K=1$ and $K=4$.
The average results over the randomly generated automata for verifying weak $K$-step opacity are depicted in Figure \ref{fig:weak_exp}.
Due to the long runtime of the $K$-delay state estimator and two-way observer in some cases, we omit these results when necessary.
As in the strong case, the forward comparison method performed nearly identically to the secret observer method.
From these plots, we see that the proposed methods for verification outperform the existing $K$-delay state estimator in average runtime and size in all cases.
While the runtime in applying the two-way observer is smaller for small-sized automata, the secret observer method outperforms it on the average in time and space for larger automata ($>15$ states).
It should be noted that one property of this method for generating random automata is that for larger system sizes, nearly all of the automata generated were opaque for each notion of $K$-step opacity.
We consider a more balanced and structured method for generation next.

\subsection{Second random generation approach (grid-based)}

In the second experiment, we generate automata as a square grid where states can transition to the 4 adjacent states.
These transitions are then randomly removed or labeled with a random event.
The number of observable events and secret states are scaled logarithmically with the system size.
Again, all states are considered initial.
The generation of these automata was tuned to provide a balance of automata that were opaque and not opaque across all system sizes.

We present results for verifying strong $K$-step opacity in Figure \ref{fig:grid_based}.
These results show similar trends to the previous method for generating random automata.
One notable difference is that the two-way observer method for verifying weak $K$-step opacity offers slightly improved performance over the proposed secret observer method.

\vspace{0.5cm}

\section{Conclusion}
\label{sec:conclusion}
We have presented several new results for the information-flow property of opacity in the context of discrete event systems.
We presented a general framework of opacity to unify the many existing notions across a variety of system and intruder models.
We used this framework to discuss notions of opacity over automata, both language-based and state-based.
We provided several methods for verification of language-based opacity.
We then developed a general approach for specifying state-based notions of opacity with automata and a transformation of these notions to language-based ones.
Together, we used these results to describe existing notions of opacity like current-state opacity and initial-state opacity.
We demonstrated how our approach unifies existing methods for opacity by showing the resulting language-based verification methods for these notions embody the existing verification methods.
We further demonstrated the effectiveness of this approach in our investigation of $K$-step and infinite step opacity.

Using the intuition of $K$-step opacity with our approach, we derived a uniform view of four notions of $K$-step and infinite-step opacity.
Two of these notions correspond to the existing notions of strong and weak $K$-step opacity, while the other two are new and meaningful notions.
We developed appropriate specification automata for these notions, allowing verification with the language-based methods.
We formally analyzed the complexity of these methods for $K$-step and infinite step opacity, showing these methods compare favorably in some instances to existing methods.
In particular, we showed that the proposed secret observer method outperforms the existing $K$-delay estimators for verifying strong and weak $K$-step opacity.
Finally, we performed numerical experiments with randomly-generated automata to compare the verification methods.
These results showed that the proposed verification methods offer increased performance over existing methods.

It would be interesting to apply our approach of specifying notions of opacity to capture more specific notions of privacy and security for real systems and evaluate the corresponding verification methods.
These notions could capture time-dependent notions of privacy like $K$-step opacity or multiple notions of privacy arranged hierarchically.
As we express opacity in a language-based way,
any method for checking regular-language containment could be used for verification.
For example, lattice-based methods as in \cite{doyenAntichainsAutomataBasedApproach2009} could be used for verification while avoiding the complexity of explicit determinization required by the methods presented here.
Additionally, it would be useful to extend the proposed framework to consider notions of opacity beyond the binary property considered here.
For example, probabilistic opacity in a stochastic setting \cite{yinInfinitestepOpacityStochastic2017}, approximate opacity for systems with numerical observations \cite{yinApproximateOpacityCyberPhysical2020}, or quantifying levels of opacity \cite{berardQuantifyingOpacity2015}.

Finally, it would be interesting to use the proposed framework for opacity in the context of enforcement.
Enforcement involves the synthesis of mechanisms to alter the system in order to guarantee opacity.
As the framework expresses state-based notions of opacity in a language-based manner, existing language-based synthesis methods could be leveraged to enforce more general notions of opacity.
For example, enforcement of opacity via supervisory control has been studied in \cite{dubreilSupervisoryControlOpacity2010a}.
Additionally, enforcement via obfuscation as in \cite{wuSynthesisObfuscationPolicies2018a, wuSynthesisInsertionFunctions2014a} appears to be readily implementable with this approach.

\bibliographystyle{spmpsci}
\bibliography{main}{}

\end{document}